\documentclass[onecolumn]{emulateapj}
\usepackage[utf8]{inputenc}
\usepackage{subfigure}
\usepackage{amsmath,amsfonts}
\usepackage{graphicx}
\usepackage{natbib}
\usepackage{changepage}

\newcommand{\kepler}{\textit{Kepler}}
\newcommand{\numax}{$\nu_{\rm max}$}
\newcommand{\flicker}{$F_8$}

\begin{document}

\title{Predicting Granulation ``Flicker" and Radial Velocity ``Jitter" from Spectroscopic Observables}
\author{Jamie Tayar}
\affiliation{Institute for Astronomy, University of Hawaii, 2680 Woodlawn Drive, Honolulu, HI 96822, USA  }
\affiliation{Ohio State University, 140 W. 18th Ave, Columbus, OH 43210, USA}
\affiliation{Hubble Fellow}
\author{Keivan G.\ Stassun}
\affiliation{Vanderbilt University, Department of Physics \& Astronomy, 6301 Stevenson Center Ln., Nashville, TN 37235, USA}
\affiliation{Fisk University, Department of Physics, 1000 17th Ave.\ N., Nashville, TN 37208, USA}
\author{Enrico Corsaro}
\affiliation{INAF--Osservatorio Astrofisico di Catania, via S.\ Sofia 78, 95123 Catania, Italy}

\begin{abstract}
Surface granulation can be predicted with the mass, metallicity, and frequency of maximum power of a star. Using the orders-of-magnitude larger APOGEE-\textit{Kepler} sample, we recalibrate the relationship fit by \citet{Corsaro:2017} for ``flicker", an easier-to-compute diagnostic of this granulation. We find that the relationship between the stellar parameters and flicker is significantly different for dwarf and subgiant stars than it is for red giants. We also confirm a dependence of flicker amplitude on metallicity as seen originally by \citet{Corsaro:2017}, although the dependence found here is somewhat weaker. Using the same APOGEE-\kepler\ sample, we demonstrate that spectroscopic
measurements alone provide sufficient information to estimate the flicker amplitude to {7 percent for giants, and 20 percent for dwarfs} and subgiants. We provide a relationship that depends on effective temperature, surface gravity, and metallicity, and calculate predicted flicker values for 129,000 stars with APOGEE spectra. Finally, we use published relationships between flicker and radial velocity jitter to estimate minimum jitter values for these same 129,000 stars, 
and we identify stars whose total jitter is likely to be even larger than the granulation-driven jitter by virtue of large-amplitude photometric variability. 

\end{abstract}

\section{Introduction}\label{sec:intro}

One of the main challenges to measuring precise masses of extrasolar planets is that many stars exhibit Doppler noise with an amplitude comparable to or larger than the radial-velocity (RV) signal of the planet \citep[e.g.,][]{Saar:1998,Wright:2005,Pont:2011}. This is especially the case for very low-mass planets and/or planets on wide orbits, where the RV amplitude of the stellar reflex motion can be on the order of $\sim$1~m~s$^{-1}$ or smaller. 
If the star is highly magnetically active, such that the stellar surface has large spots, the RV noise can be tens of m~s$^{-1}$ in amplitude \citep[e.g.,][]{Dumusque:2011}, though it may also be sufficiently coherent that it can be modeled and removed, in particular if a simultaneous light curve is available \citep[see, e.g.][]{Aigrain:2012,Dumusque:2014}. 

RV noise can also arise from less coherent phenomena such that it is much more challenging to mitigate. For example, stellar plages represent a kind of chromospheric activity-related phenomenon in solar-type stars that can produce significant RV variations (so-called ``RV jitter") that may be less amenable to modeling from the light curve. Nonetheless, it may be possible for some dwarf stars to estimate the amplitude of the plage-driven RV noise if a measure of the stellar rotation period is available, such as from a light curve \citep[see, e.g.,][]{Saar:2003}. More generally, attempts have been made to establish empirical relationships between chromospheric activity proxies (e.g., UV emission, Ca $R'_{HK}$, etc) and RV jitter \citep[e.g.,][]{Cegla:2014}. 

However, RV jitter can also arise---at levels of up to $\sim$20~m~s$^{-1}$---in stars that are otherwise devoid of magnetic activity-driven variability \citep[e.g.,][]{Wright:2005}, including very slowly rotating dwarfs, subgiants, and red giant stars. Indeed, RV planet surveys have been stymied in some cases from achieving their nominal RV precision of $\lesssim$1~m~s$^{-1}$, despite carefully selecting stars known to be chromospherically inactive \citep[see, e.g.,][]{Wright:2005,Isaacson:2010} and mitigating instrumental and astrophysical effects \citep{LovisFischer2010, BeattyGaudi2015}. 
In the era of large-scale searches for small, transiting planets via precise light curves, such as {\it CoRoT\/} \citep{Auvergne:2009}, {\it MOST\/} \citep{Walker:2003}, {\it Kepler\/} \citep{Borucki:2010}, and {\it TESS\/} \citep{Ricker:2015}, it has become increasingly important to evaluate individual stars for the possibility of high RV jitter that may not be otherwise discernible from the light curve itself or from other activity proxies. For example, a key science goal for {\it TESS\/} is to identify at least 50 Earth-like transiting planets whose masses can be measured from follow-up RV observations \citep{Ricker:2015}.


A likely driver of RV jitter that is independent of magnetic activity is convective motions at the stellar surface (i.e., granulation), as demonstrated by, e.g., \citet{Bastien:2014}. 
Indeed, \citet{Bastien:2014} found that, among otherwise photometrically ``quiet" stars (whose overall photometric variability is $\lesssim$3 parts per thousand; ppt), the observed RV jitter can range in amplitude from $<$4~m~s$^{-1}$ to 20~m~s$^{-1}$, and that this RV jitter is most strongly correlated with the granulation ``flicker". 
The 8-hour Flicker ($F_8$) first defined by \citet{Bastien:2013} refers to the low-level photometric variations arising from granulation in the integrated light of a star occurring on timescales shorter than 8~hr. The typical $F_8$ amplitudes in white light (such as in a {\it Kepler\/} light curve) range from $\approx$0.015~ppt to $\approx$0.4~ppt, depending most directly on the stellar surface gravity, $\log g$ \citep{Bastien:2016}.
Because the $F_8$ amplitude is so small, it generally can only be measured in an ultra-precise light curve such as from {\it Kepler\/} or {\it TESS}, but it provides access to predicting RV jitter amplitudes for otherwise photometrically ``quiet" stars whose RV noise would be difficult to predict in other ways. 

More recently, \citet{Oshagh:2017} demonstrated that the predictive power of $F_8$ extends to stars with overall photometric activity amplitudes as large as $\sim$10~ppt and RV jitter as high as $\sim$100~m~s$^{-1}$. 
Thus, measuring or estimating $F_8$ for large numbers of stars to be targeted by {\it TESS\/} could serve to refine the RV followup strategy by focusing on those planet candidates whose stellar hosts are most likely to be sufficiently RV stable to enable precise measurement of planetary masses. 
At the same time, through its strong correlation with $\log g$, $F_8$ provides an opportunity to measure stellar masses and radii for stars that will be observed by {\it TESS\/} and {\it Gaia\/} \citep[see, e.g.,][]{Stassun:2018}. Including both {\it TESS\/} targets slated for 2-min cadence observations \citep{Stassun_TIC:2018} as well as those that will be observed in the 30-min cadence full-frame images \citep[e.g.,][]{Oelkers_FFI:2018}, should enable precise, fundamental stellar masses and radii for tens of thousands of stars \citep[see][]{Stassun:2018}. 

Recent work has also demonstrated that while $F_8$ correlates most fundamentally with $\log g$, opacity effects cause it to also depend on the stellar metallicity. For example, \citet{Corsaro:2017} used a small sample of red giant stars in open clusters observed by {\it Kepler\/} to show that a change in [M/H] of $\sim$0.5~dex can produce a change in $F_8$ of $\sim$50\%. Accounting for the metallicity dependence of $F_8$ can therefore improve predictions of granulation-driven RV jitter and also improve estimates of stellar $\log g$. Indeed, \citet{Stassun:2018} found in a sample of {\it Kepler\/} stars having {\it Gaia\/} parallaxes and asteroseismically determined masses and radii \citep[e.g.,][]{Huber:2017}, the precision on the stellar $\log g$ inferred from $F_8$ improved from $\sim$0.1~dex to $\sim$0.05~dex when accounting for the dependence of $F_8$ on [M/H]. Consequently, the precision with which individual stellar masses can be estimated with $F_8$ from a {\it TESS\/} light curve improves from $\sim$20\% to $<$10\% \citep[see][]{Stassun:2018}. 
In turn, such precise stellar measurements will permit even more precise determinations of planet masses and radii \citep[see, e.g.,][]{Johns:2018,Berger:2018,Stassun:2017}. 

The purpose of this paper is twofold. First, we seek to use the large sample of stars with {\it Kepler\/} light curves for which accurate masses, radii, temperatures, and metallicities are now also available in order to refine the relationships between $F_8$, $\log g$, and [M/H]. Second, we aim to use these refined relationships to predict the granulation-driven RV jitter amplitudes for a large number of stars that will be observed by {\it TESS}. 

In Sec.~\ref{sec:data} we summarize the data that we use in this study, as well as the method that we use to robustly estimate the dependence of $F_8$ on $\log g$, [M/H], and other parameters. 
Sec.~\ref{sec:results} presents the main results of this study, including our refined $F_8$ relationships, their application to a very large sample of likely {\it TESS\/} targets having spectroscopic parameters in order to infer their $F_8$, and finally our predicted RV jitter amplitudes for this large sample of stars. 
We discuss these results, caveats, and guidelines for their use in Sec.~\ref{sec:discussion}. 
Finally, we conclude with a summary in Sec.~\ref{sec:summary}.

\section{Data}\label{sec:data}
In this work, we use two spectroscopically characterized samples of stars: a smaller calibration sample with asteroseismic measurements, and a broader sample with only spectroscopic information to which we apply the relationship we derive. We discuss the spectroscopic and asteroseismic characterization of the 2465 stars from the APOGEE-\kepler\ overlap sample used for calibration in
Section \ref{ssec:apokasc} and the broader sample of 129,055 stars with only APOGEE spectra in Section \ref{ssec:apogee}. For the calibration sample, we also require \flicker\ measurements from \kepler, which are described in Section \ref{ssec:flicker}. For the larger sample, it is important to identify active stars likely to deviate from the flicker relation, and our method for doing so is discussed in Section \ref{ssec:kelt}. 

\subsection{Spectroscopy \& Asteroseismology}

\subsubsection{The APOGEE-\kepler\ sample} \label{ssec:apokasc}




In order to calibrate {an accurate} relationship between $F_8$, RV jitter, and spectroscopic observables, we choose to use the combined APOGEE-\textit{Kepler} (APOKASC) sample of dwarfs, subgiants, and first ascent red giants. This sample is ideal for this sort of work because it has a large number of stars covering a wide range for parameter space. They have been precisely characterized using both asteroseismology, the study of stellar oscillations (e.g. \citealt{Aerts2010}), and high resolution spectroscopy. Because of the change in frequency of oscillations as a star moves across the HR diagram, the asteroseismic characterization of these stars is naturally broken up into a dwarf/subgiant sample, which oscillates more rapidly and requires short cadence data, and a red giant sample which can be characterized using the standard long cadence data from the \kepler\ mission. 

Parameters for dwarfs and subgiants are taken from the analysis of \citet{Serenelli:2017}. These stars represent a subset of the \citet{Chaplin2011} sample of stars with detected seismic oscillations in the short cadence \textit{Kepler} data. Seismic analysis was carried out using several pipelines, with the central values taken from the SYD pipeline \citep{Huber2009}, and improved via comparison with grids of stellar models. 
Spectroscopic parameters for these stars are taken from Data Release 14 \citep[DR14,][]{DR14} of APOGEE-2 \citep{Majewski2017}, one of the components of the Sloan Digital Sky Survey IV \citep[SDSS-IV,][] {Blanton2017}, which is using the 2.5m Sloan Digital Sky Survey telescope \citep{Gunn2006} to take H-band spectra. The observations are normalized and compared to a grid of synthetic spectra by the automated ASPCAP pipeline \citep{Nidever2015, GarciaPerez2016} and stellar parameters are determined by a global chi-squared minimization. For consistency with the red giant seismic analysis, we use the dwarf and subgiant results determined using the APOGEE spectroscopic temperatures. 

We add to the dwarf and subgiant sample the sample of first ascent red giants analyzed in \citet{Pinsonneault2018} by five different seismic pipelines. The resulting seismic parameters were corrected for pipeline dependent systematics, averaged, calibrated to an absolute scale using the open clusters in the \textit{Kepler} field, and corrected for deviations from homology \citep{Serenelli:2017}. This analysis also uses temperatures from APOGEE DR14 \citep{DR14}. We use only stars determined to be first ascent red giants by consensus of the APOKASC collaboration \citep[][see also Y. Elsworth et al. in prep]{Pinsonneault2018} to avoid additional correlations in the expected flicker signal with evolutionary state \citep{Bastien:2016} 
and because most red clump stars fall outside of the validated gravity range of the flicker technique. We have also excluded KIC 9893440, which has a close companion contaminating the light curve, and thus an anomalous flicker measurement for its surface gravity. To prevent other unidentified binary stars from contaminating our results, we have also required that the stars in our sample have radial velocity variability of less than 100 m~s$^{-1}$ in the APOGEE spectra. 

\begin{figure}[!ht]
\centering
\includegraphics[width=0.495\linewidth]{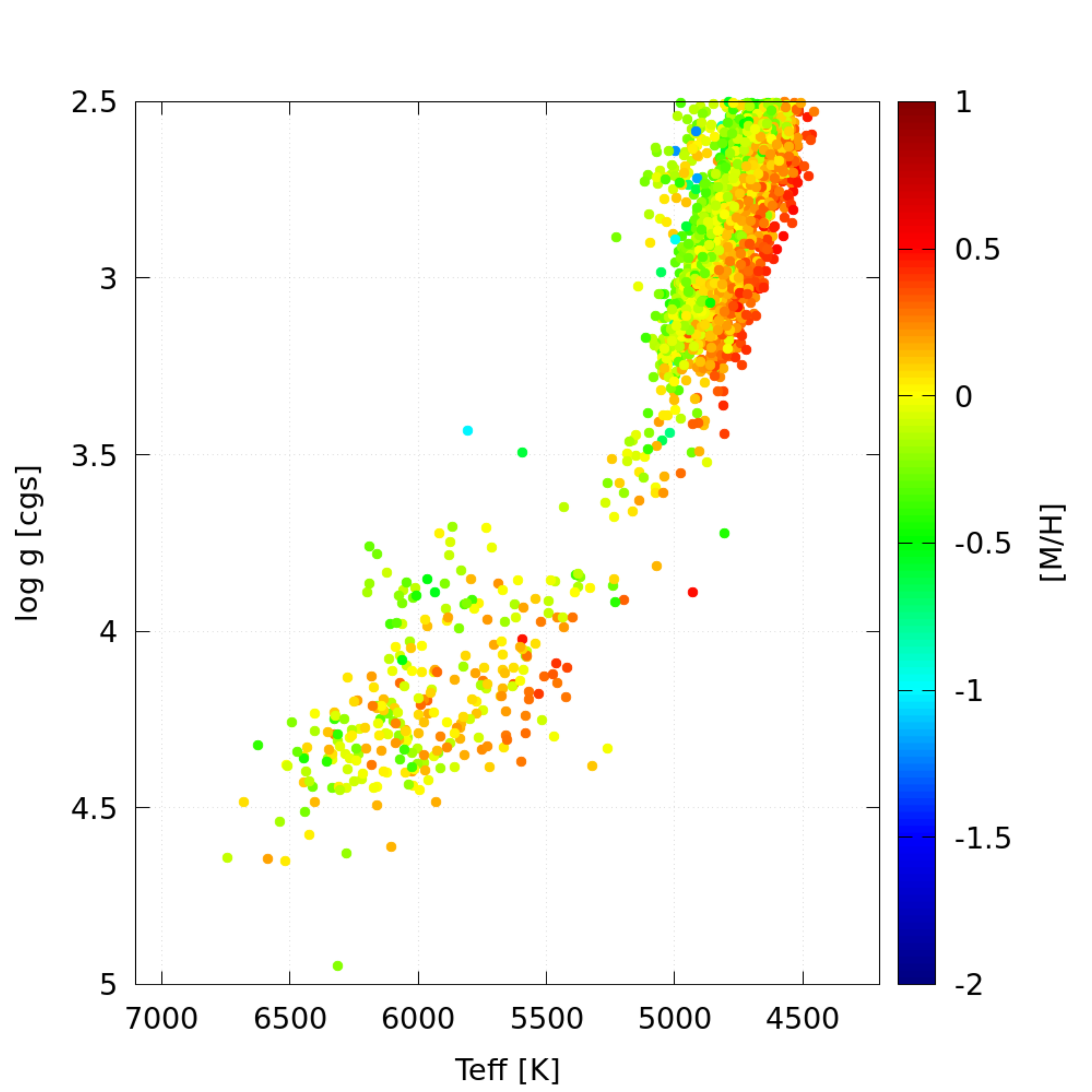}
\includegraphics[width=0.495\linewidth]{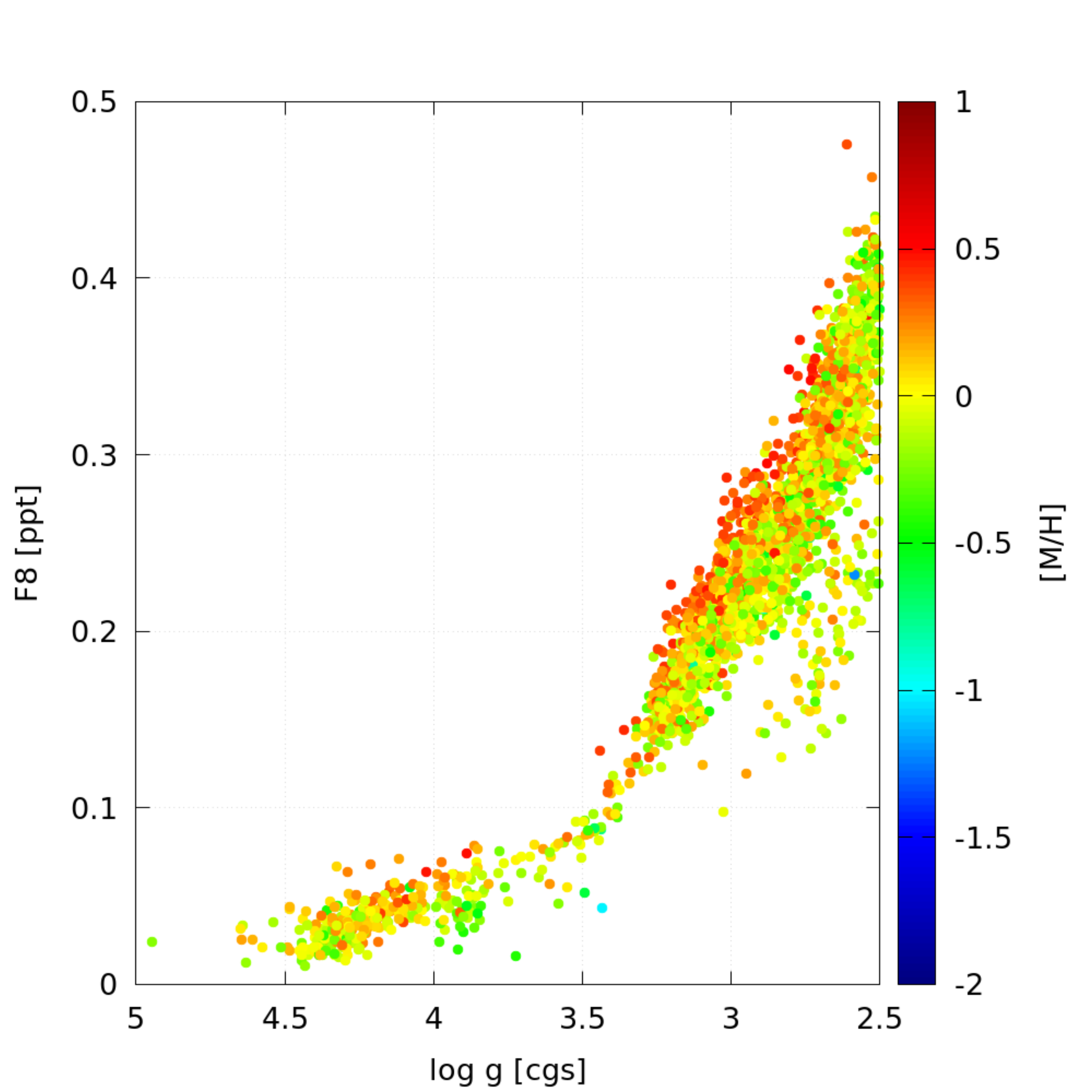}
\caption{ Left: Kiel diagram of the APOKASC stars used in this analysis, color coded by metallicity. Right: The dependence of flicker on surface gravity, with the color coding indicating a secondary dependence on metallicity. }
\label{Fig:APOKASC}
\end{figure}

Additionally, we restrict both our dwarf/subgiant and giant samples to stars where the flicker ($F_8$) is considered a reliable proxy of surface gravity via granulation,  specifically, stars with log(g) between 4.6 and 2.5 dex \citep{Bastien:2016}.
We indicate the temperatures and gravities of the stars in our APOKASC sample in Figure \ref{Fig:APOKASC}. Basic data for all the APOKASC stars used in this analysis, including their evolutionary state, and listed in Table \ref{Tab:seismic}. 
\\

\begin{table}[htbp]

\caption{Basic data for the 2465 stars in our calibration sample. Spectroscopic information comes from the APOGEE survey \citep{DR14}, asteroseismic values come from \citet{Pinsonneault2018} for giants and \citet{Serenelli:2017} for dwarfs and subgiants, and flicker information is calculated here using the formula of \citet{Bastien:2016}. The full table is provided in the electronic version of the journal. A portion is shown here for guidance regarding form and content.}
\begin{adjustwidth}{-0.1cm}{} 
\tiny
\begin{tabular}{rlrrrrrrrrrrrrrl}
\hline
\hline
\multicolumn{1}{l}{KIC ID} & 2MASS ID & \multicolumn{1}{l}{\numax} & \multicolumn{1}{l}{$\sigma_{\nu max}$} & \multicolumn{1}{l}{Mass} & \multicolumn{1}{l}{$\sigma_{Mass}$} & \multicolumn{1}{l}{[Fe/H]} & \multicolumn{1}{l}{$\sigma_{[Fe/H]}$} & \multicolumn{1}{l}{T$_{\rm eff}$} & \multicolumn{1}{l}{$\sigma_{T_{\rm eff}}$} & \multicolumn{1}{l}{log (g)} & \multicolumn{1}{l}{$\sigma_{log (g)}$} & \multicolumn{1}{l}{$F_8$} & \multicolumn{1}{l}{$\sigma_{F_8}$} & \multicolumn{1}{l}{$v_{sc}$} & Ev. St. \\ 
& & $(\mu$Hz) & ($\mu$Hz) & (M$_\sun$) & (M$_\sun$) & (dex) & (dex) & (K) & (K) & (cgs) & (cgs) & (ppt) & (ppt) & (kms) &  \\
\hline
10000207 & 2M19052985+4654372 & 94.6 & 0.9 & 1.04 & 0.04 & -0.18 & 0.03 & 4703 & 74 & 2.79 & 0.06 & 0.27 & 0.01 & 0 & RGB \\ 
10000547 & 2M19062193+4657016 & 160.5 & 1.4 & 1.14 & 0.05 & -0.20 & 0.04 & 4968 & 85 & 3.11 & 0.06 & 0.20 & 0.01 & 0 & RGB \\ 
10001326 & 2M19081378+4655345 & 47.6 & 0.4 & 1.20 & 0.05 & -0.02 & 0.03 & 4656 & 73 & 2.57 & 0.06 & 0.39 & 0.03 & 0 & RGB \\ 
10003270 & 2M19122841+4658290 & 781.1 & 54.1 & 1.31 & 0.07 & -0.17 & 0.13 & 6199 & 175 & 3.89 & 0.10 & 0.05 & 0.00 & 0 & D/S \\ 
10003349 & 2M19123799+4656309 & 48.8 & 0.6 & 1.36 & 0.06 & 0.02 & 0.03 & 4687 & 75 & 2.56 & 0.06 & 0.32 & 0.02 & 0 & RGB \\ 
10004825 & 2M19153149+4657298 & 57.5 & 0.5 & 1.66 & 0.07 & 0.25 & 0.03 & 4690 & 77 & 2.71 & 0.05 & 0.29 & 0.01 & 0 & RGB \\ 
10006097 & 2M19175990+4654261 & 140.6 & 1.3 & 1.11 & 0.05 & 0.04 & 0.03 & 4855 & 84 & 3.02 & 0.06 & 0.23 & 0.01 & 0 & RGB \\ 
10007492 & 2M19202078+4656427 & 214.9 & 1.9 & 1.46 & 0.06 & 0.00 & 0.03 & 4994 & 90 & 3.18 & 0.06 & 0.15 & 0.00 & 0 & RGB \\ 
10014959 & 2M19325640+4654288 & 174.2 & 1.6 & 1.05 & 0.04 & 0.15 & 0.03 & 4740 & 72 & 3.13 & 0.05 & 0.22 & 0.01 & 0 & RGB \\ \hline

\end{tabular}
\label{Tab:seismic}
\end{adjustwidth}
\end{table}


\subsubsection{The APOGEE sample} \label{ssec:apogee}




Once we have fit a relation that predicts the flicker and the radial velocity jitter as a function of spectroscopic observables, we want {to extend those predictions to a larger sample of stars.} 
For this {purpose}, we take the stars from the full APOGEE survey, specifically Data Release 14 \citep{DR14}. 
Where possible, we use corrected APOGEE parameters, otherwise we use the raw values from the ASPCAP pipeline.


\subsection{Light Curves and Variability Measures}

\subsubsection{Flicker from {\it Kepler}} \label{ssec:flicker}
In order to develop our updated relationships between granulation Flicker ($F_8$) and other stellar parameters, we require a set of calibration stars for which precise $F_8$ measurements are available. The most extensive $F_8$ measurements published are those of \citet{Bastien:2016}, which provides $F_8$-based $\log g$ determinations for 27,628 {\it Kepler\/} stars brighter than 13.5~mag with 4500~K $< T_{\rm eff} <$ 7150~K, $2.5 < \log g  < 4.6$, and overall photometric amplitudes of less than 10~ppt. 
The $F_8$ values in the catalog published by \citet{Bastien:2016} were erroneous \citep[see also][]{Oshagh:2017}, and to our knowledge the corrected values have not been published. Therefore, we redetermined the $F_8$ values for the 2465 stars in our calibration sample using the same {\it Kepler\/} light curves and employing the same methodology as in \citet{Bastien:2016}.

\subsubsection{Photometric Variability from {\it KELT}} \label{ssec:kelt}
Previous work examining the ability of $F_8$ to predict RV jitter has found that the \flicker-jitter correlation holds for stars that are photometrically ``quiet", presumably because the RV jitter becomes increasingly dominated by magnetic activity effects (e.g., spots, plage) which manifest as large amplitude photometric variations. In particular, \citet{Bastien:2014} estimated that the \flicker-jitter relationship holds for stars with overall light curve amplitudes less than $\sim$3~ppt. \citet{Oshagh:2017} found that the \flicker-jitter relationship continues to hold for stars with overall light amplitudes as high as $\sim$10~ppt. 

{\it TESS\/} will observe stars over nearly the entire sky, so to estimate the photometric variability amplitudes of as many stars as possible, we use the variability catalog of 4~million stars observed by the Kilodegree Extremely Little Telescope \citep[KELT;][]{Pepper:2007,Pepper:2012} survey \citep{Oelkers:2018}. 
In our analysis and results we flag stars whose photometric variability amplitudes have been determined to be less than the thresholds mentioned above. 

\section{Methods \& Results}\label{sec:results}

\subsection{Recalibrating the Asteroseismic Flicker Relation}




As discussed in \citet{Bastien:2016}, the flicker signal depends most strongly on surface gravity, and this is consistent with previous work that indicates that stellar granulation, for which flicker is a proxy, correlates with \numax\ \citep{Hekker2012}. More recent analysis by \citet{Corsaro:2017} has shown that for red giants, stellar granulation also depends on the mass of the star and on its metallicity. However, the fit by \citet{Corsaro:2017} (here after C17) was done using only 60 stars over the relatively small metallicity range, covered by the open clusters NGC 6811, NGC 6819 and NGC 6791, namely $-0.09 < \mbox{[M/H]} < 0.32$.

\begin{figure}[!ht]
\centering
\includegraphics[width=.31\textwidth]{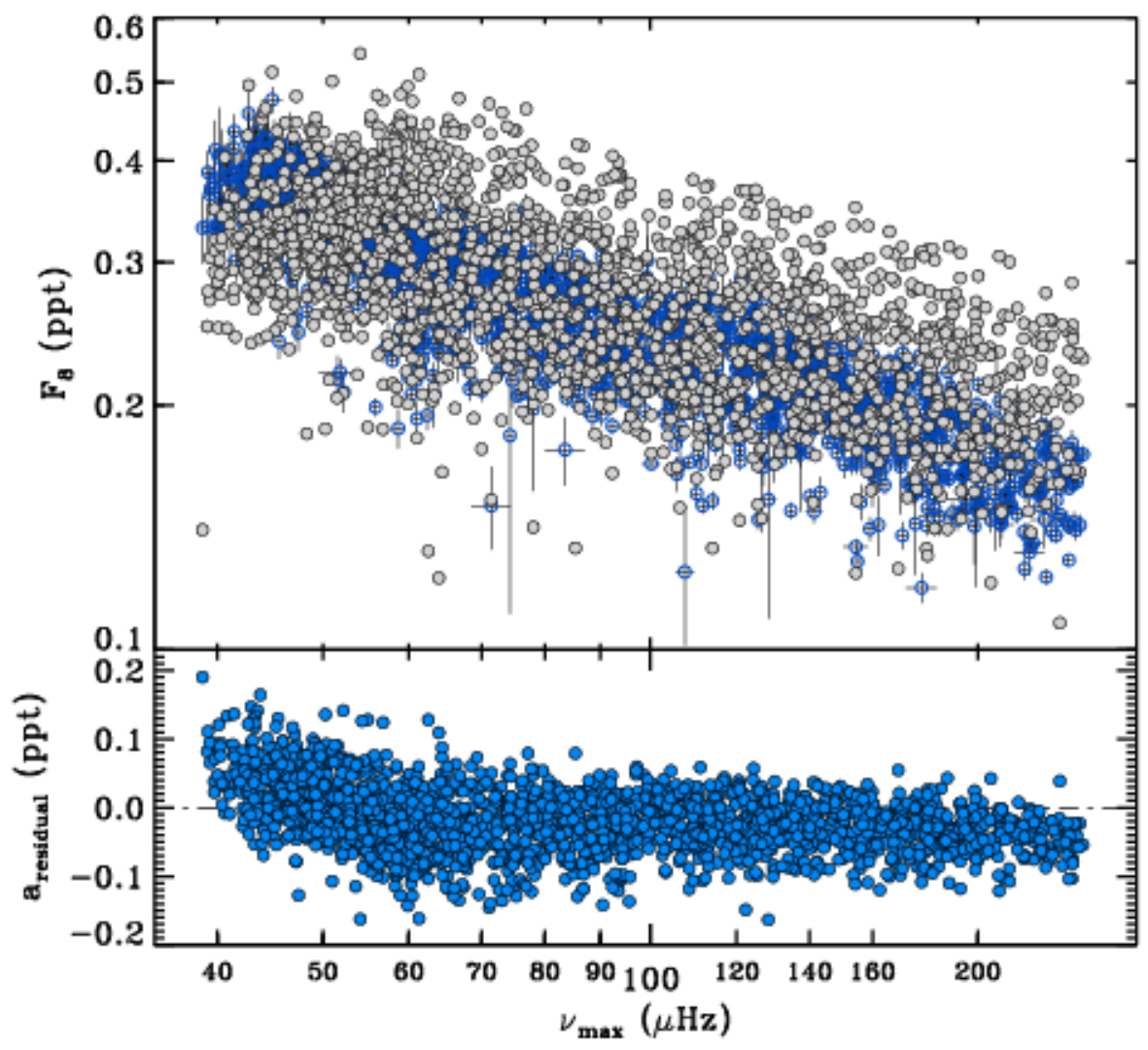}\includegraphics[width=.31\textwidth]{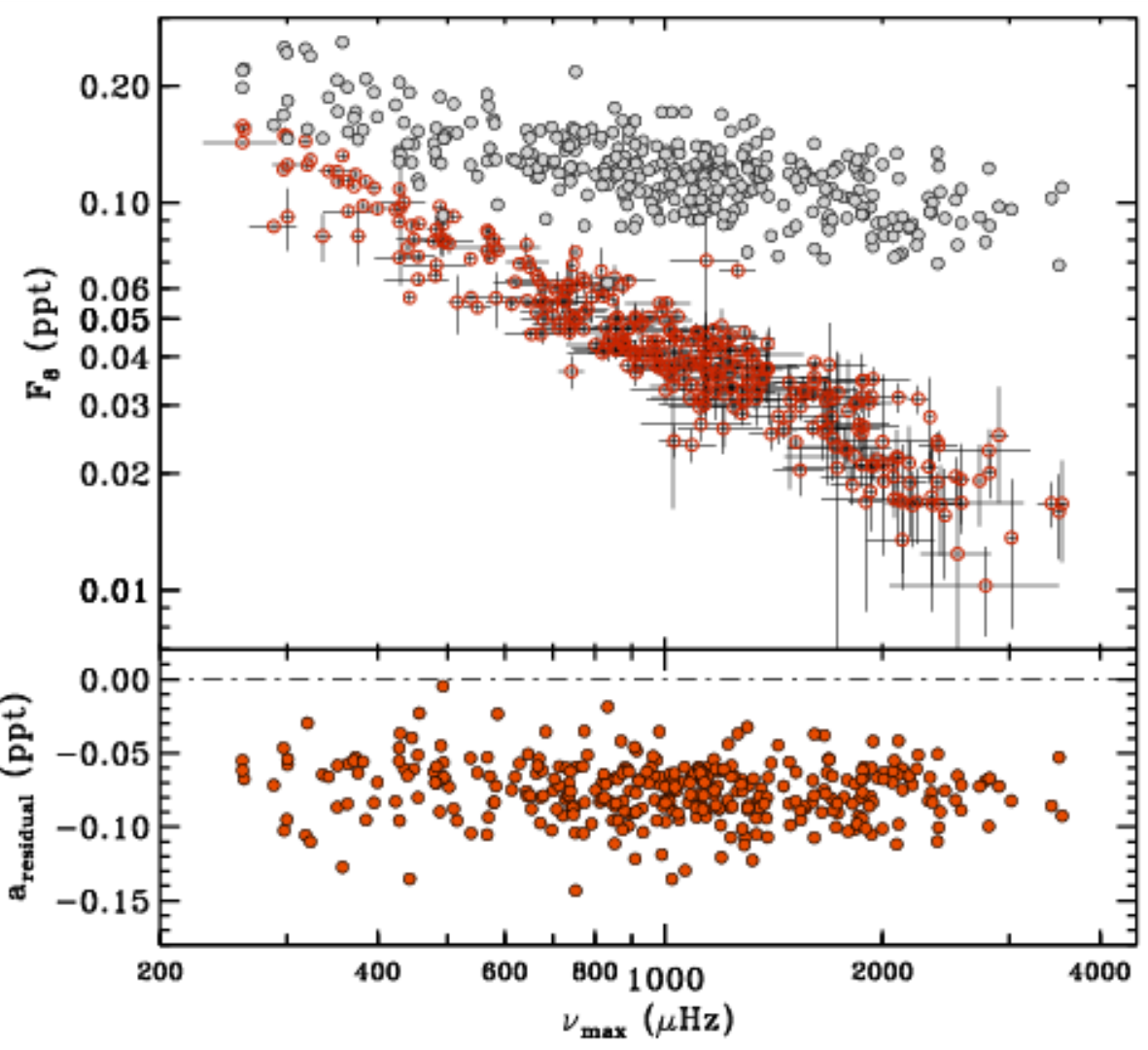}

\includegraphics[width=.31\textwidth]{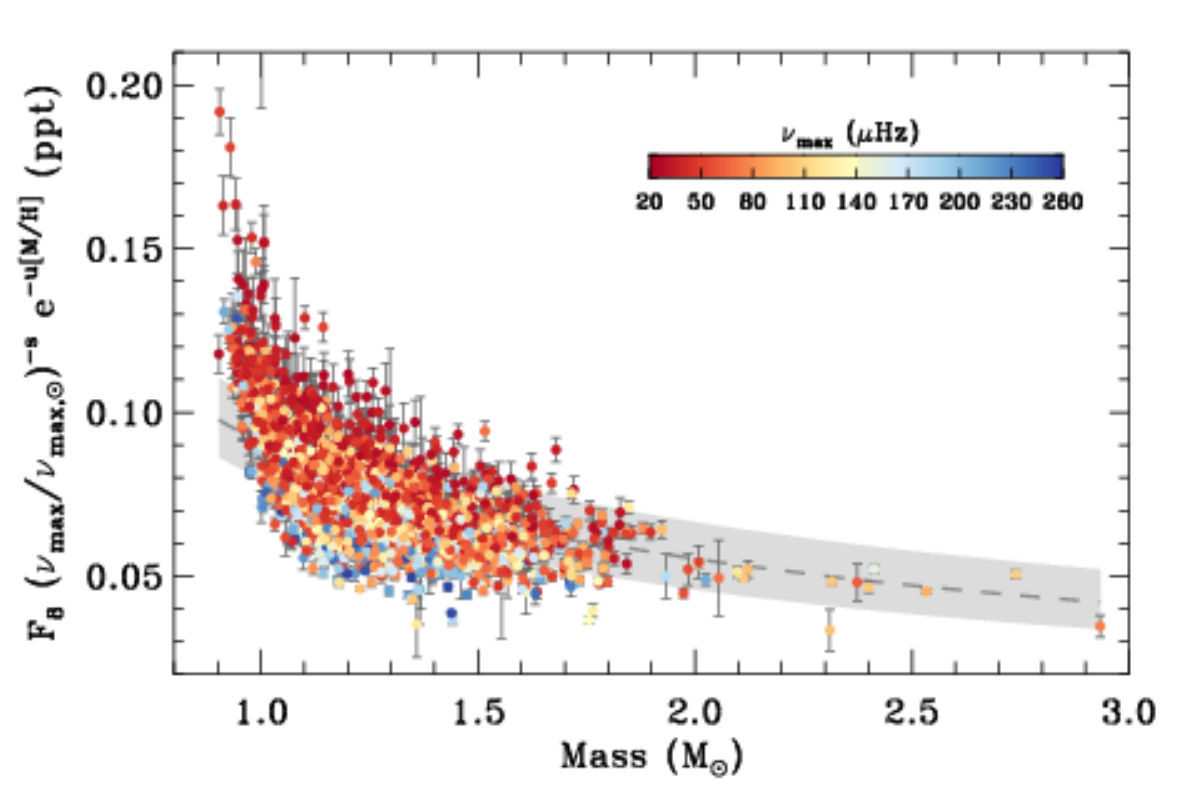}\subfigure{\includegraphics[width=.31\textwidth]{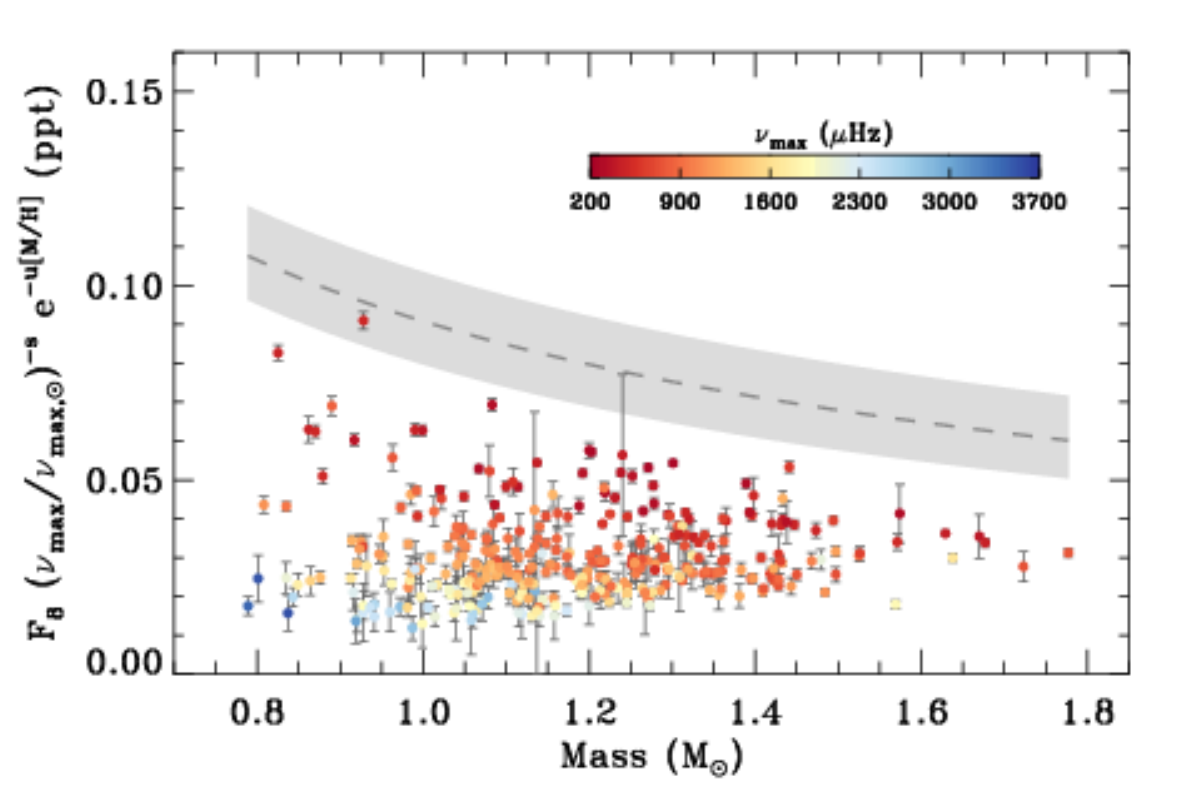}}

\includegraphics[width=.31\textwidth]{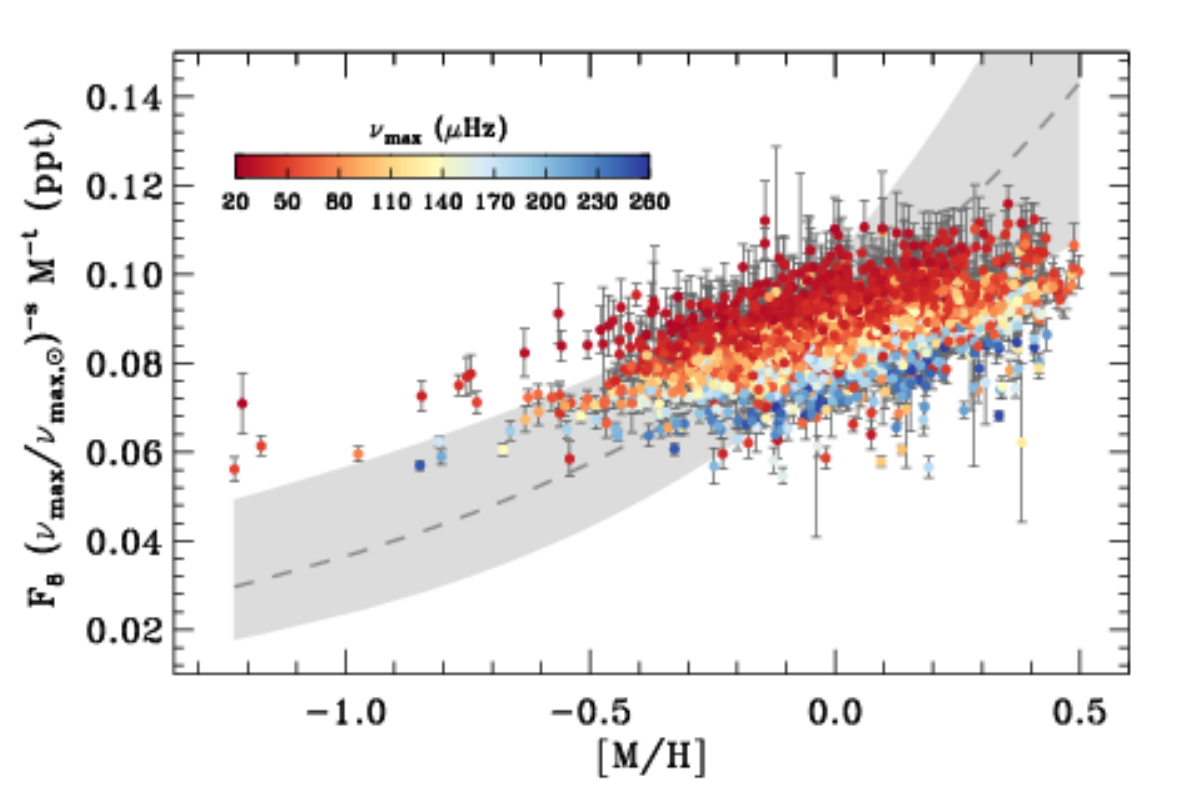}\includegraphics[width=.31\textwidth]{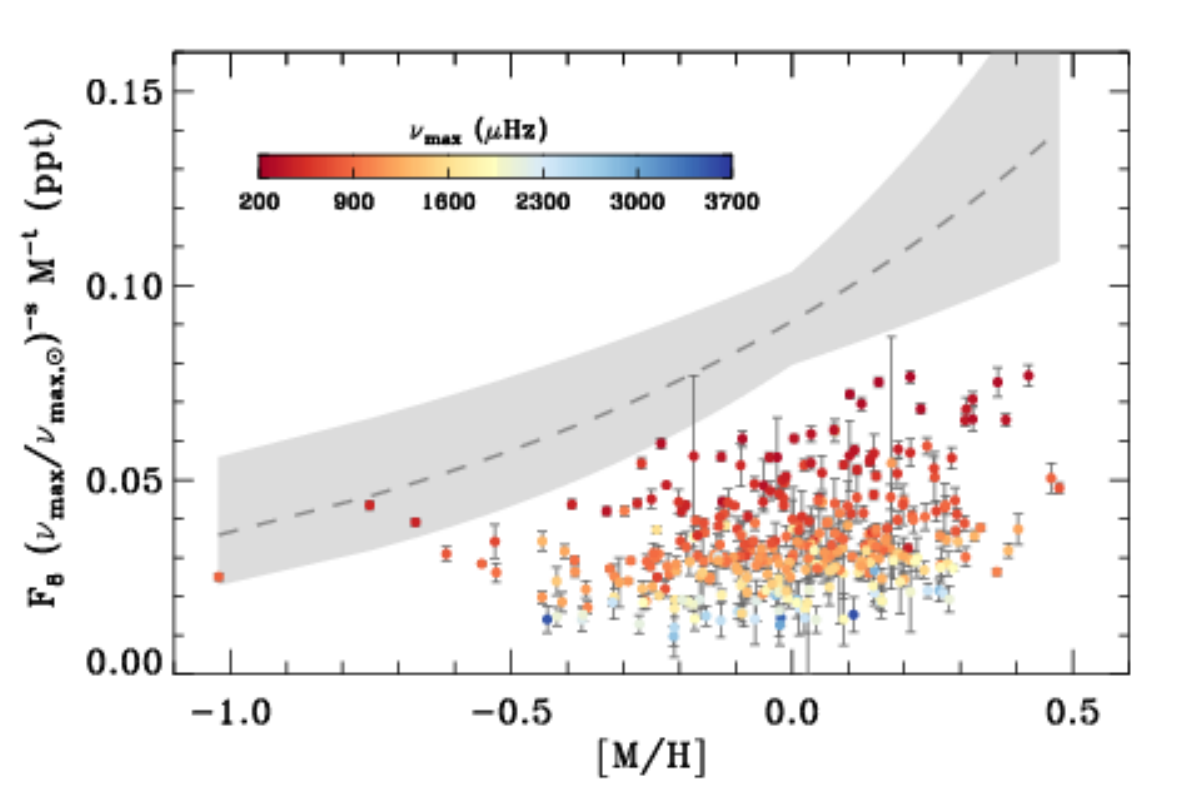}

\caption{Comparisons between the predictions from the original C17 relationships focused on the \numax, mass, and metallicity dependence (from top to bottom) for both the LC red giant stars (left panels) and the SC dwarfs and subgiants (right panels). Bayesian credible intervals of 68.3\,\% are shown with gray shaded regions. Note that to make the comparison possible, the original parameter ranges of mass, metallicity and \numax\, used by C17 have been extended to match the parameter ranges of our sample of stars.}
\label{Fig:C17}
\end{figure}

We show in Figure \ref{Fig:C17} that when the C17 relation is applied to the full metallicity and gravity range of our APOKASC sample, the resulting predictions are not consistent with the observed $F_8$ values. This is particularly true in the dwarf regime, a result that is not surprising since the C17 relationship was not calibrated in this range of surface gravities. In general, the C17 predictions for dwarfs are overestimated with a rather constant offset across the range of surface gravities. In addition, in the case of giant stars, the C17 relation implies a metallicity dependence that seems somewhat overestimated compared with observations.

{Given that we have measurements of $F_8$ for our APOKASC sample, accounting for over an order of magnitude more stars than the original C17 paper, and the difference arising from the C17 predictions as highlighted in Figure~\ref{Fig:C17}, we perform a new fit using the same Bayesian framework as C17 in order to quantify the differences arising when using our new sample of stars.} Specifically, we use the scaling law originally introduced by C17 (their Eq. 17)
\begin{equation}
    F_8^{\mathrm{(seismic)}} = \beta \left( \frac{\nu_{max}}{\nu_{max,\odot}} \right)^s \left( \frac{M}{M_\odot} \right)^t e^{u\mathrm{[M/H]}}
    \label{eq:F8_astero}
\end{equation}
{where \numax\ is the frequency of maximum oscillation power and directly proportional to the stellar surface gravity \citep{Brown1991}, $M$ is the stellar mass, and [M/H] the metallicity. This sort of scaling relation relies on the assumption of homology among stars, which is generally reasonable across our sample range. Estimating the exponents $(s, t, u)$ and the proportionality term $\beta$ of the scaling relation will allow us performing a direct comparison with the results obtained by \cite{Corsaro:2017} using our sample of stars. Also, following an approach similar to that used by \cite{Corsaro:2013}, we decided to calibrate this relation separately for short cadence (SC, dwarf and subgiant) and long cadence (LC, giant) samples because, as already shown in Figure~\ref{Fig:C17}, the two regimes exhibit significantly different dependencies on the parameters considered.}

\begin{table}
\centering
 \caption{Median values of the inferred parameters $(s, t, u, \ln \beta)$ for both short cadence {(343 stars)} and long cadence {(2122 stars)} data, with the physical parameter they relate to also indicated in brackets, as well as the original fit parameters from \citet{Corsaro:2017} {(26 stars)} for comparison. Bayesian credible intervals of 68.3\,\% are included.}
\begin{tabular}{lrrrr}
\hline\hline
\\[-8pt]
Model & \multicolumn{1}{c}{$s\,(\nu_\mathrm{max})$} & \multicolumn{1}{c}{$t\,(\mbox{Mass})$} & \multicolumn{1}{c}{$u$ ([M/H])} & \multicolumn{1}{c}{$\ln \beta$}\\[1pt]
 \hline
 \\[-8pt]       
  $\mathrm{SC}$ &  $-0.885^{+0.013}_{-0.014}$ & $-0.673^{+0.048}_{-0.047}$ & $0.637^{+0.031}_{-0.034}$ & $-4.039^{+0.018}_{-0.019}$\\ [1pt]
  $\mathrm{LC}$ & $-0.467^{+0.004}_{-0.010}$ & $-0.682^{+0.005}_{-0.001}$ & $0.306^{+0.007}_{-0.008}$ & $-2.899^{+0.004}_{-0.006}$\\ [1pt]
  $\mathrm{C17}$ & $-0.350^{+0.029}_{-0.029}$ & $-0.716^{+0.077}_{-0.081}$ & $0.911^{+0.310}_{-0.307}$ & $-2.399^{+0.131}_{-0.131}$\\ [1pt]
  \hline
 \end{tabular}
\label{tab:resultsSeis}
\end{table}

\begin{figure}[!ht]
\centering
\includegraphics[width=.31\textwidth]{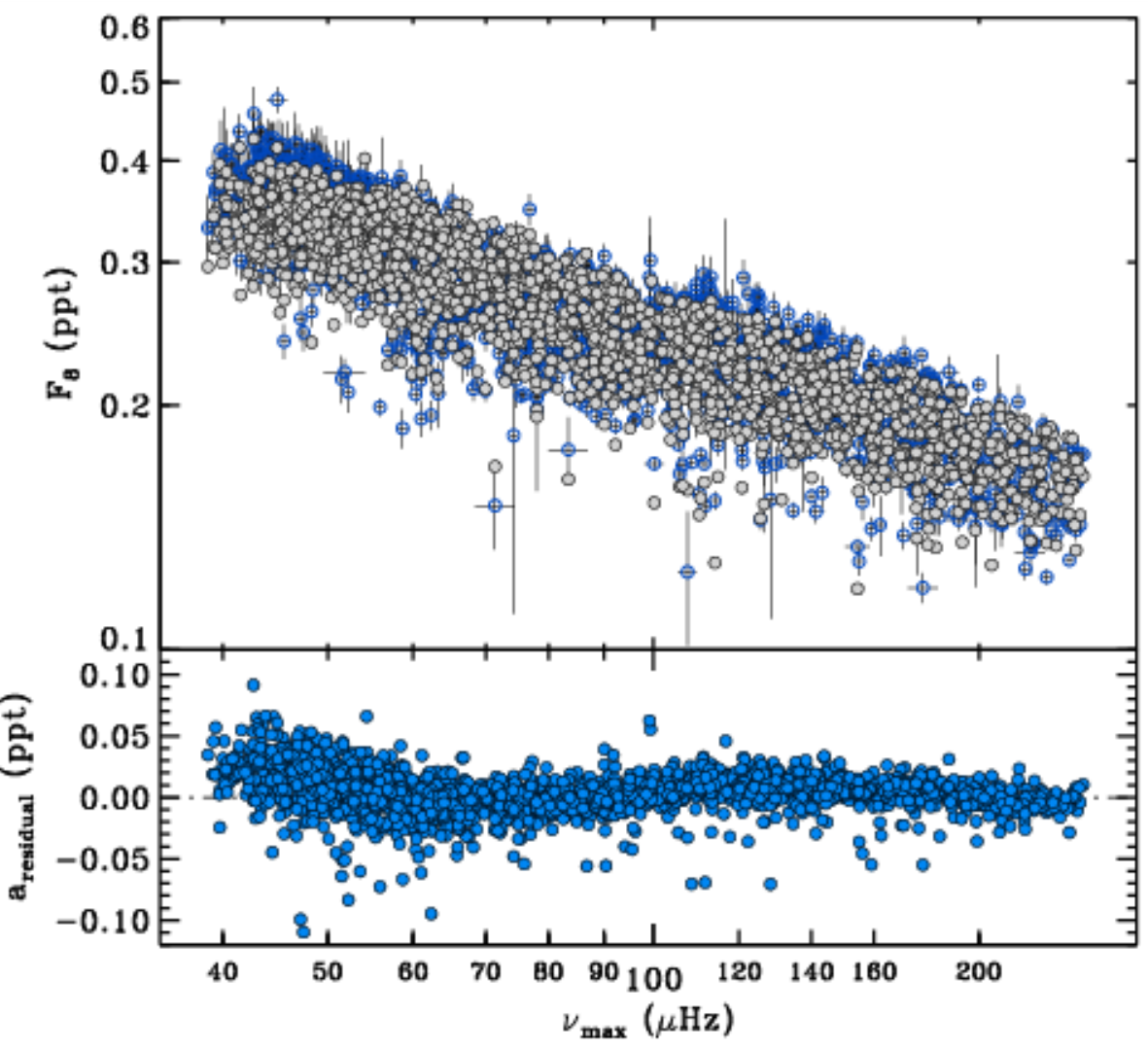}\includegraphics[width=.31\textwidth]{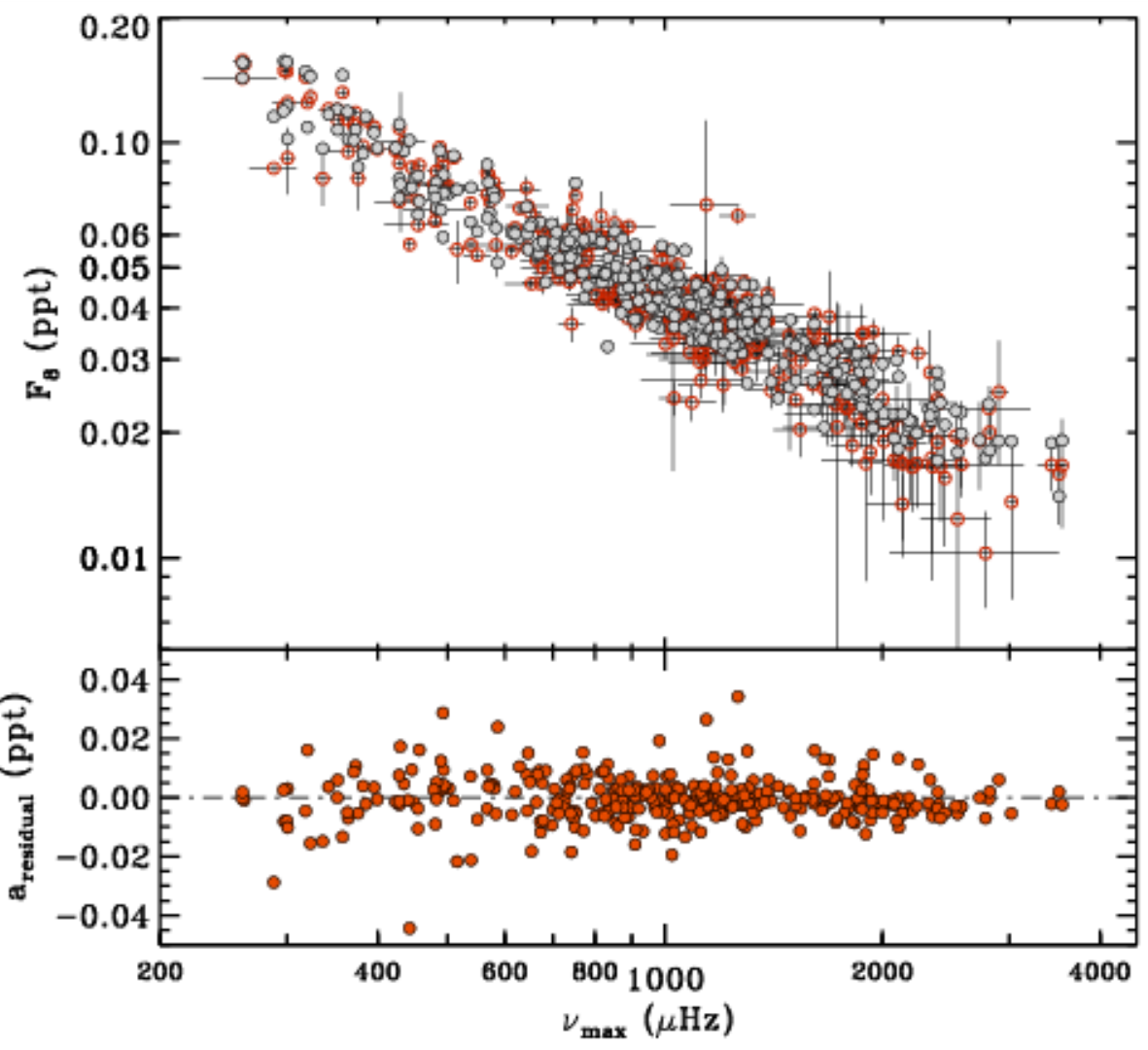}

\includegraphics[width=.31\textwidth]{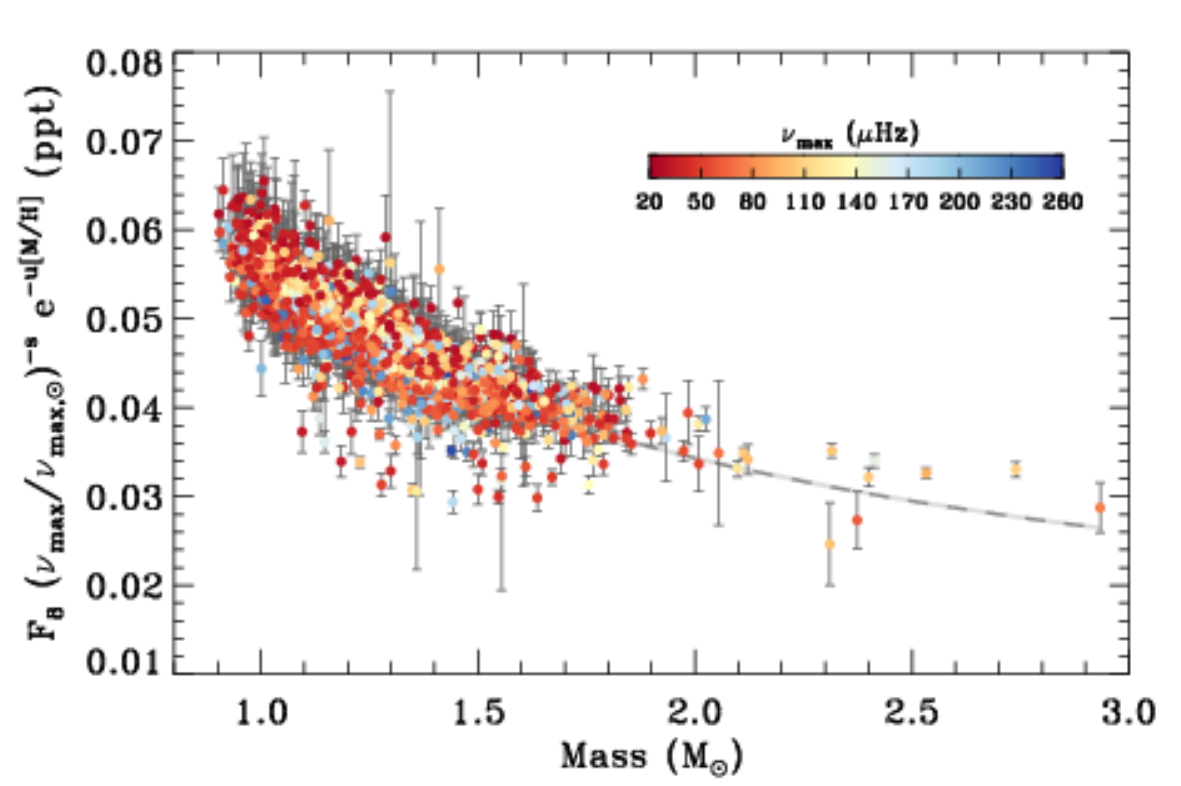}\includegraphics[width=.31\textwidth]{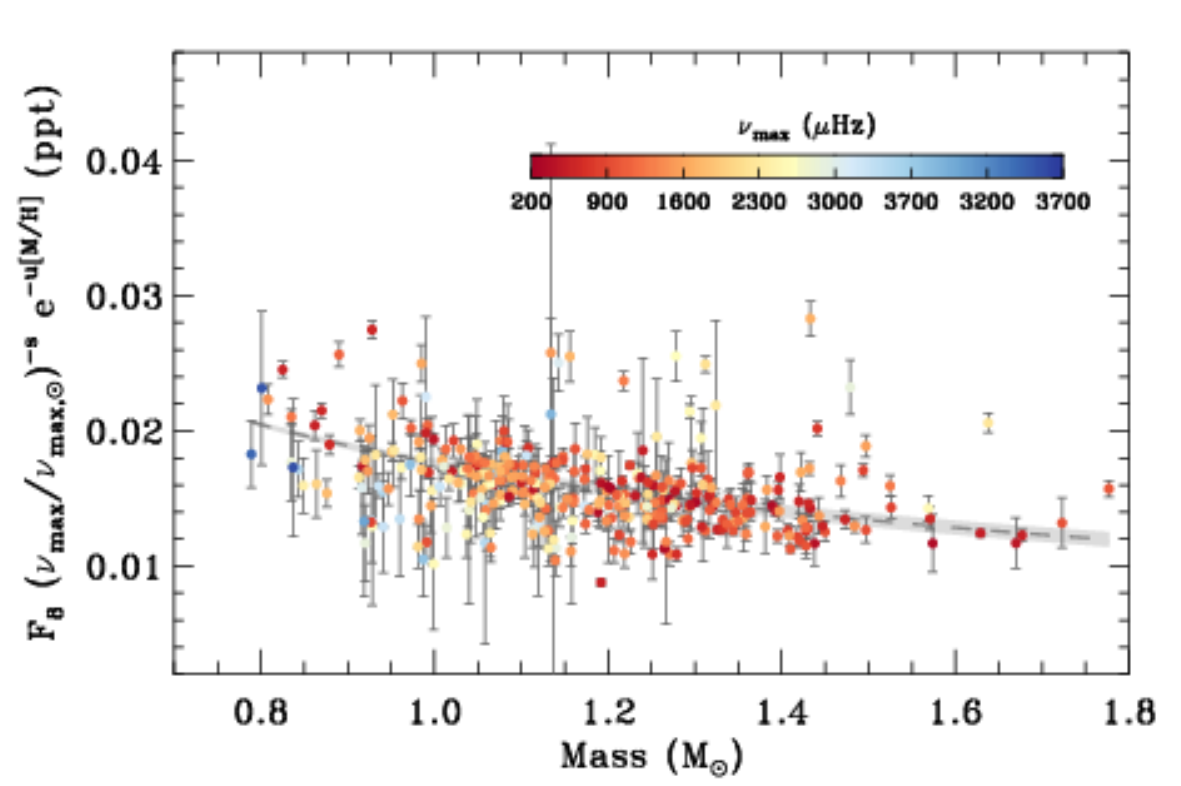}

\includegraphics[width=.31\textwidth]{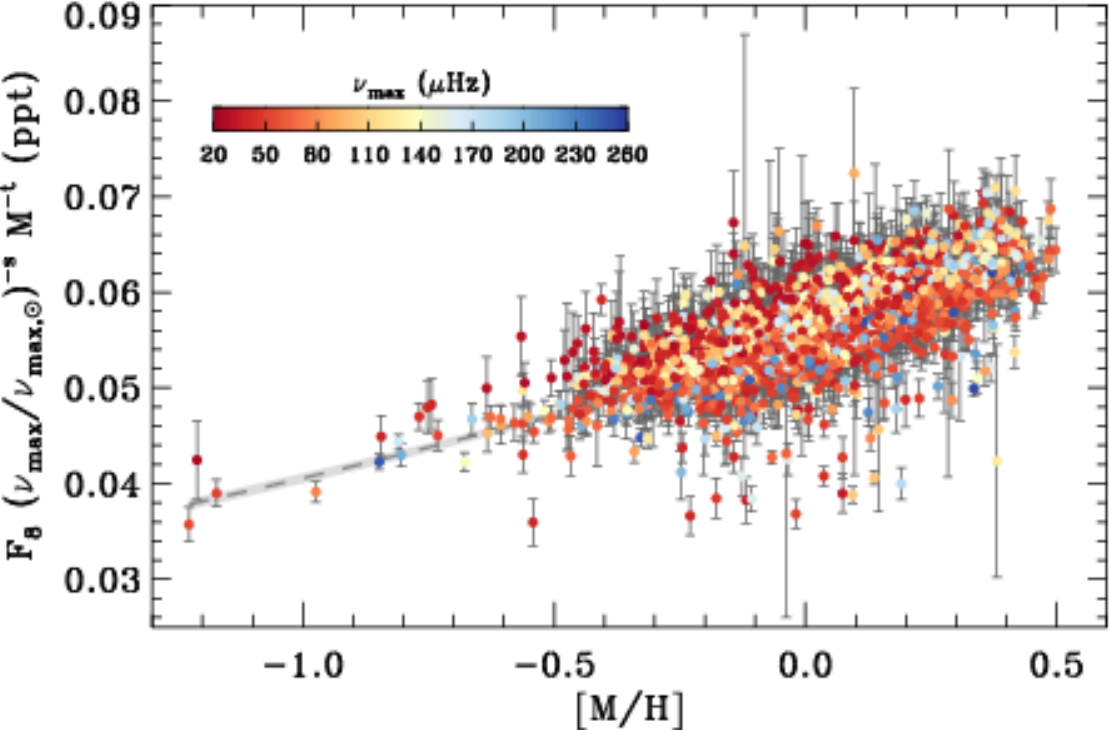}\includegraphics[width=.31\textwidth]{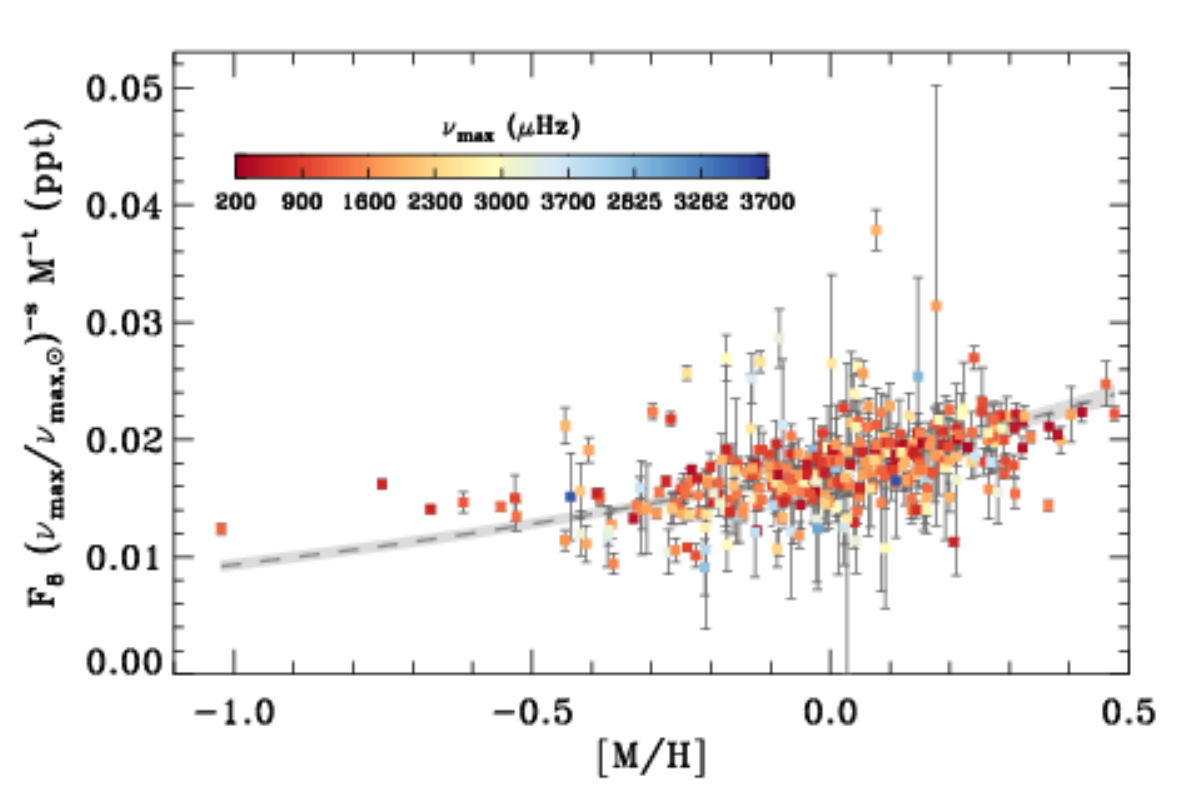}
\caption{The results of our updated fit of the dependence of flicker on asteroseismic observables. As in Figure \ref{Fig:C17} we show plots focused on the 
\numax, mass, and metallicity dependence (from top to bottom) for both the LC red giant stars (left panels) and the SC dwarfs and subgiants (right panels). Bayesian credible intervals of 68.3\,\% are shown with gray shaded regions.}
\label{Fig:newseis}
\end{figure}

{The description of how the fits are obtained is given in Sect.~\ref{sec:Bayes}. As shown in Table~\ref{tab:resultsSeis}, using Eq.~(\ref{eq:F8_astero})} we find strong evidence that the dependence of the fit on \numax\ is not the same for dwarf/subgiant sample and the giant sample, the latter being similar to the C17 result, while the former is almost twice as strong (See Figure \ref{Fig:newseis}). We find that the dependence on mass is similar for the dwarf/subgiant and giant samples, and that it is {compatible within 1-$\sigma$ to} the C17 fit. We also see the inverted trend or anti-correlation between the dependence on mass and the dependence on metallicity that was found by C17 in both LC and SC data.

{However, while interestingly the metallicity dependence for the SC sample (u $\sim$ 0.64) is compatible within the quoted error bars with that found for the cluster data ($u \sim 0.9$), the metallicity dependence of the long cadence sample is a factor of about three times smaller ($u \sim 0.3$). 
Some of this difference can be explained by the change in the average metallicity measured for stars in NGC 6791 between APOGEE Data Release 13 \citep{DR13}, used by C17, and the DR14 data used here, but additional effects of $\alpha$-element abundances and [C/N] variations in stellar granulation that were not taken into account in this analysis may also play a role. We refer the interested reader to S. Mathur et al. (in prep) for a detailed analysis and discussion of other parameters besides those in our fit that could impact the relationship investigated here. The errors on the fitted parameters of the scaling relation are significantly smaller for the new fits, especially the LC sample, than those from C17 because of the much larger number of stars used in our analysis (about two orders of magnitude more than C17 for the LC sample and about one order of magnitude for the SC sample).} For completeness, we also note that our results for the dwarf stars are consistent with the metallicity dependence found by \citet{Serenelli:2017}, although we prefer to adopt the C17 formalism because it additionally takes into account dependencies on mass and \numax.

\subsection{Calibrating the Spectroscopic Flicker Relation}
Having calibrated the relationship between precise stellar properties  of \numax, mass, and flicker, we want to determine whether spectroscopic observables alone provide sufficient information to predict the flicker value. Using the insight from the seismic sample, we again divide our stars into a dwarf/subgiant and a giant sample, using a cut at $\log g =3.4$~dex, approximately equivalent to the cut used to divide the seismic samples. We also enforce a similar form for the flicker relation, searching for the exponents $(a, b, c)$ such that 
\begin{equation}
    F_8^{\mathrm{(spec)}} = \alpha  e^{a \; \mathrm{  \log g}} \left( \frac{T_\mathrm{eff}}{T_\mathrm{eff,\odot}} \right)^b e^{c\mathrm{[M/H]}}
    \label{eq:F8_spec}
\end{equation}
{where $\log g$ is the stellar surface gravity, $T_\mathrm{eff}$ its temperature, and [M/H] is again the metallicity. $(a, b, c)$ are the corresponding exponents of the scaling relation, and $\alpha$ is a proportionality term; these are the parameters that need to be calibrated. Similarly to what done for Eq.~(\ref{eq:F8_astero}), we apply this relation separately for SC and LC observations. We expect the differences between the SC and LC fits to be even more significant here since the parameter acting as the evolutionary coordinate is log(g) in the LC sample whereas for subgiants in the SC sample $T_\mathrm{eff}$ is serving that purpose.

We deliberately choose to adopt different letters for the exponents in the two scaling relations considered, because they represent different physical relationships even though the functional forms look similar. We therefore do not expect a priori that the exponents of the two scaling laws end up in having similar estimates, even for the case of the metallicity term, which formally appears in the same way in both relations of Eqs.~(\ref{eq:F8_astero}) and (\ref{eq:F8_spec}). This is because different correlations among the observables \numax, $M$, and [M/H] on one side and $\log g$, $T_\mathrm{eff}$, and [M/H] on the other could be present, thus changing the way each term in the scaling relation will contribute to the overall fit.}

\begin{figure}[!ht]
\centering
\includegraphics[width=.31\textwidth]{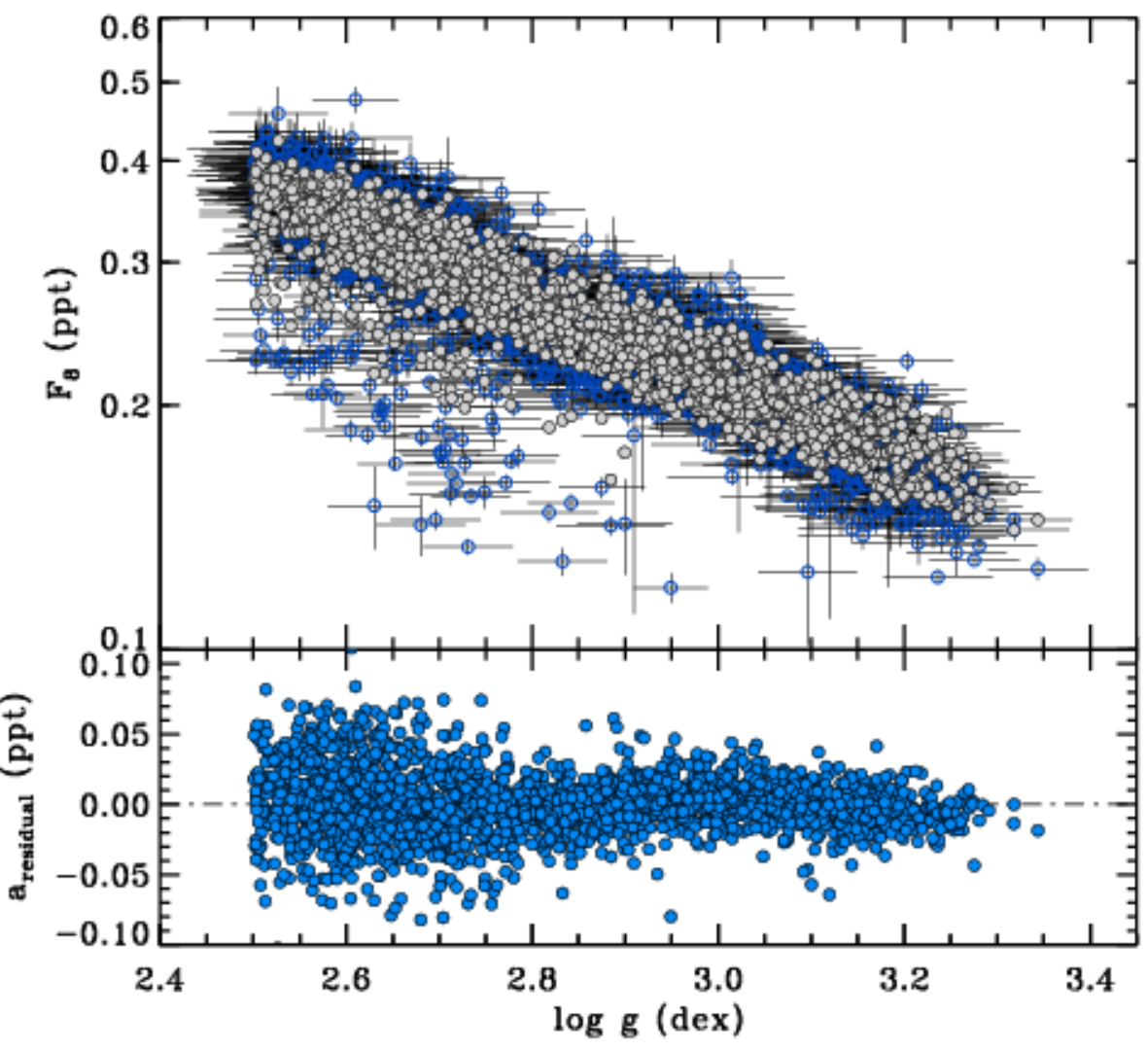}\includegraphics[width=.31\textwidth]{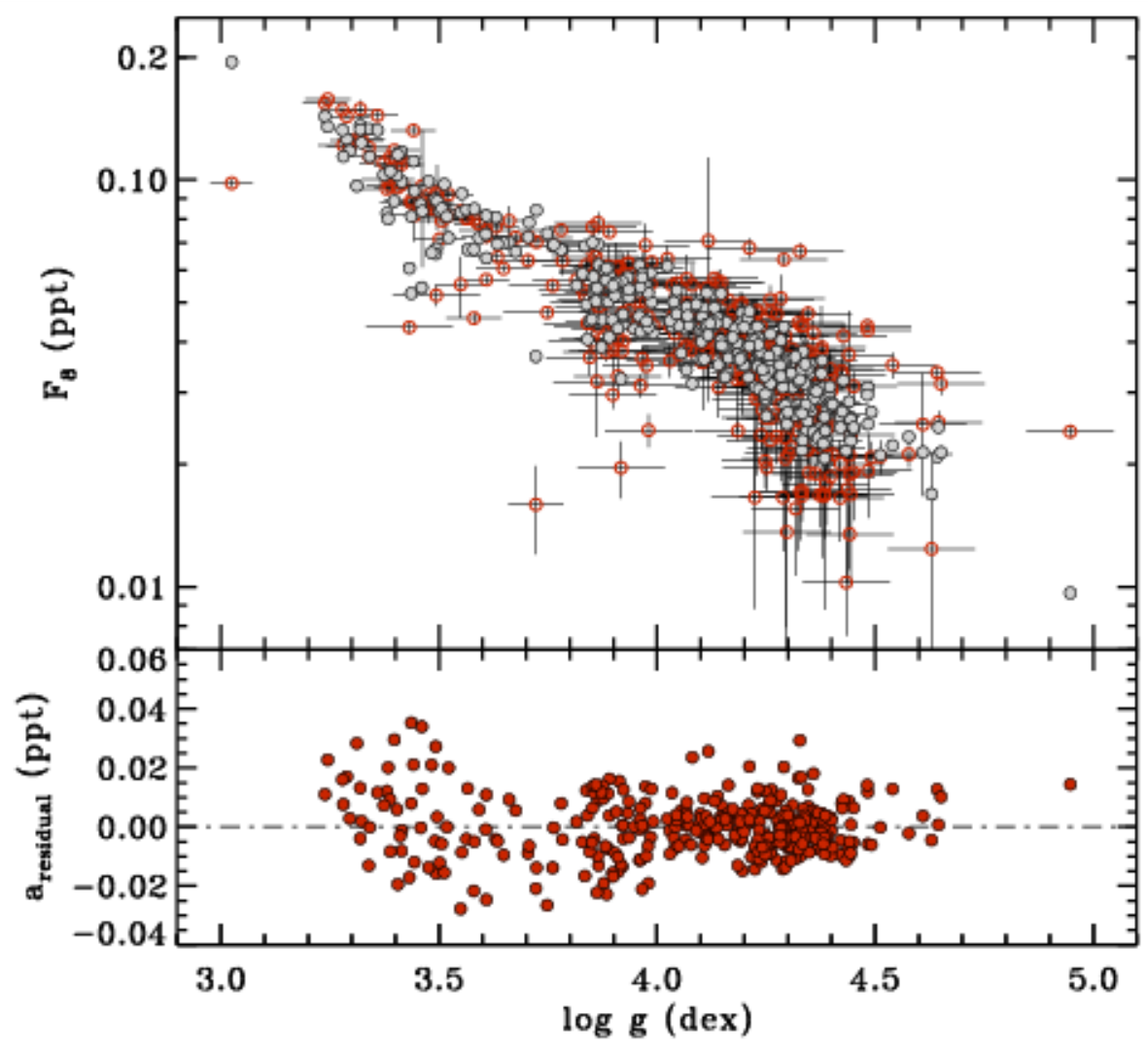}

\includegraphics[width=.31\textwidth]{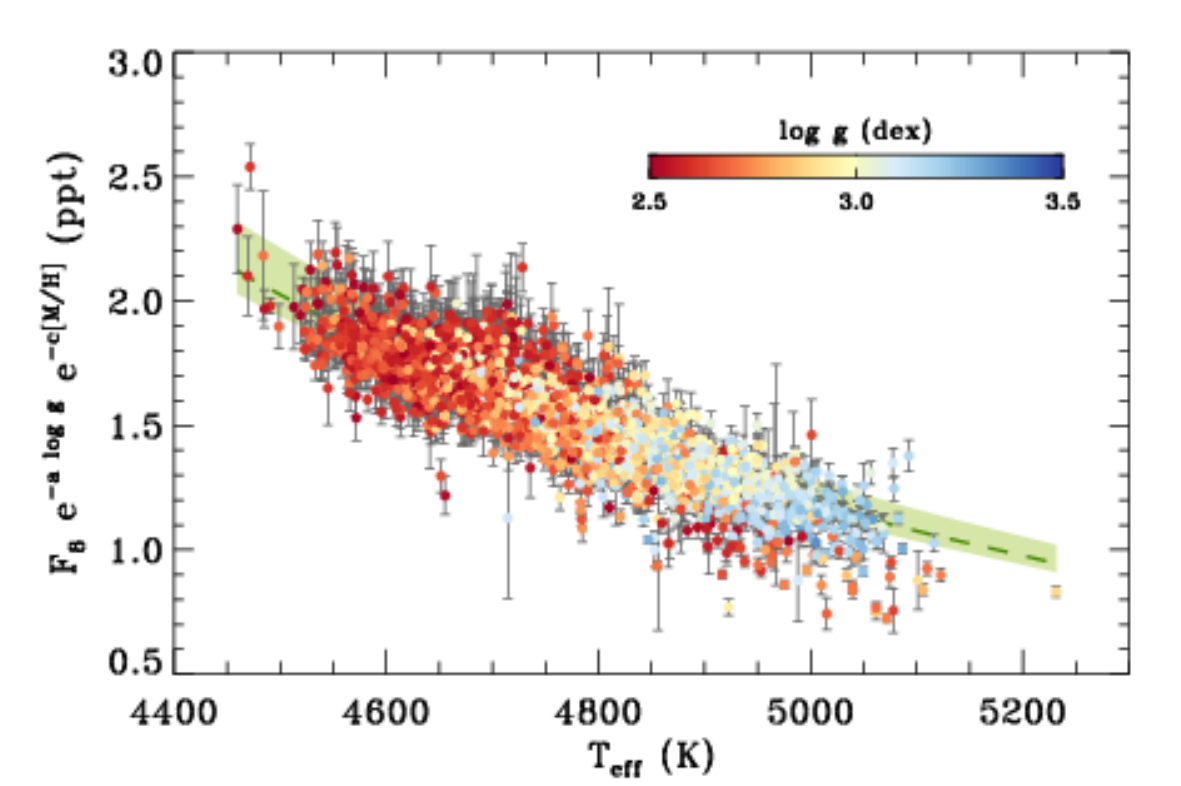}\includegraphics[width=.31\textwidth]{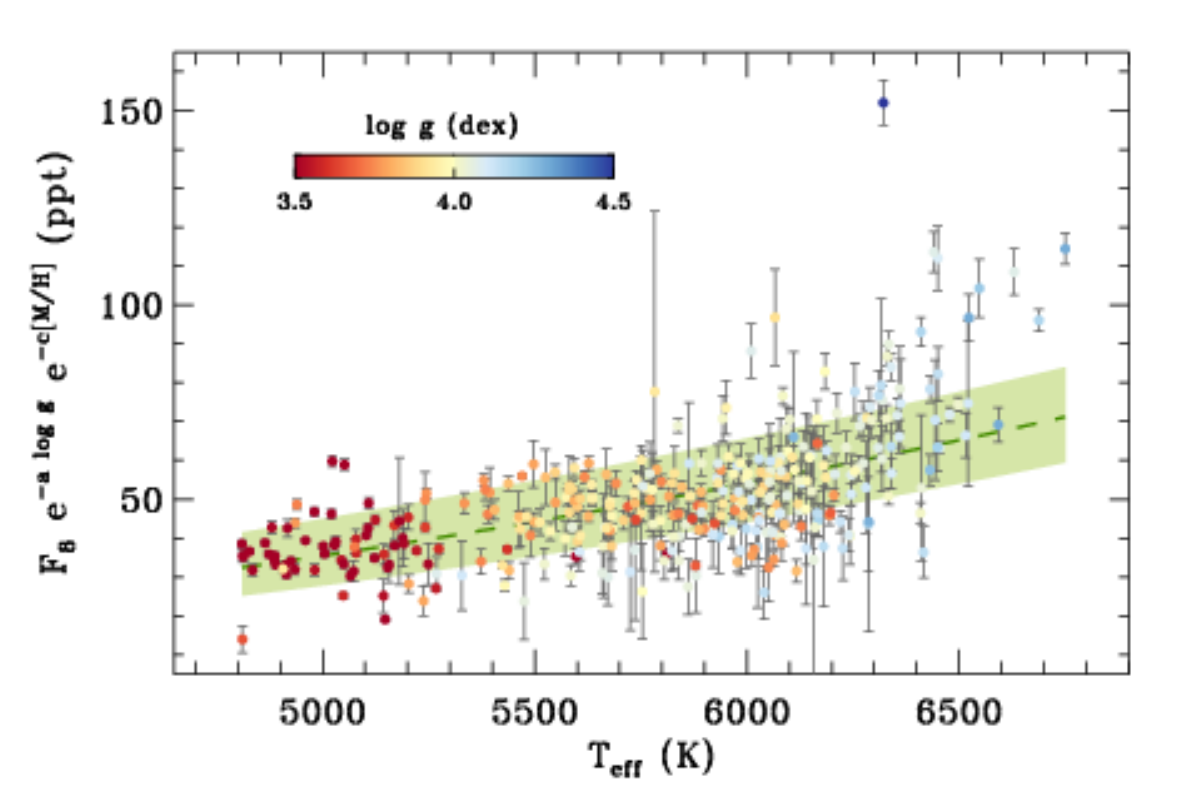}

\includegraphics[width=.31\textwidth]{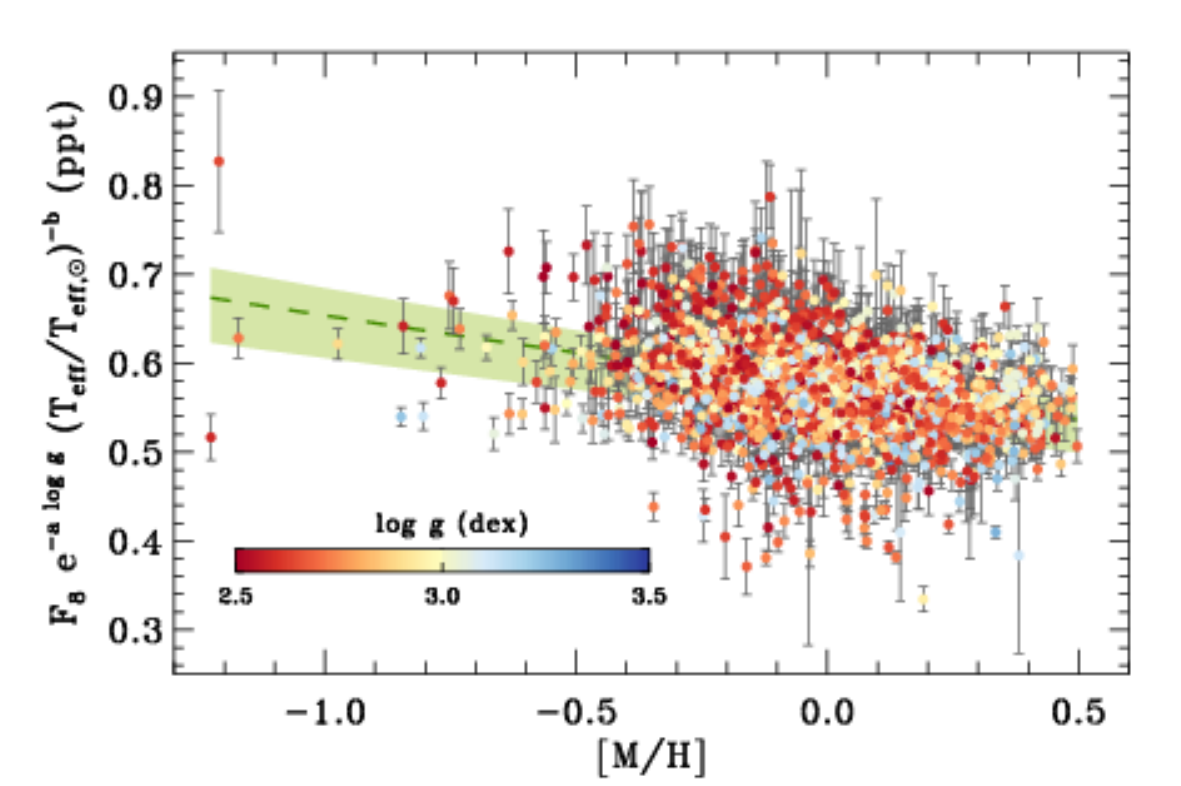}\includegraphics[width=.31\textwidth]{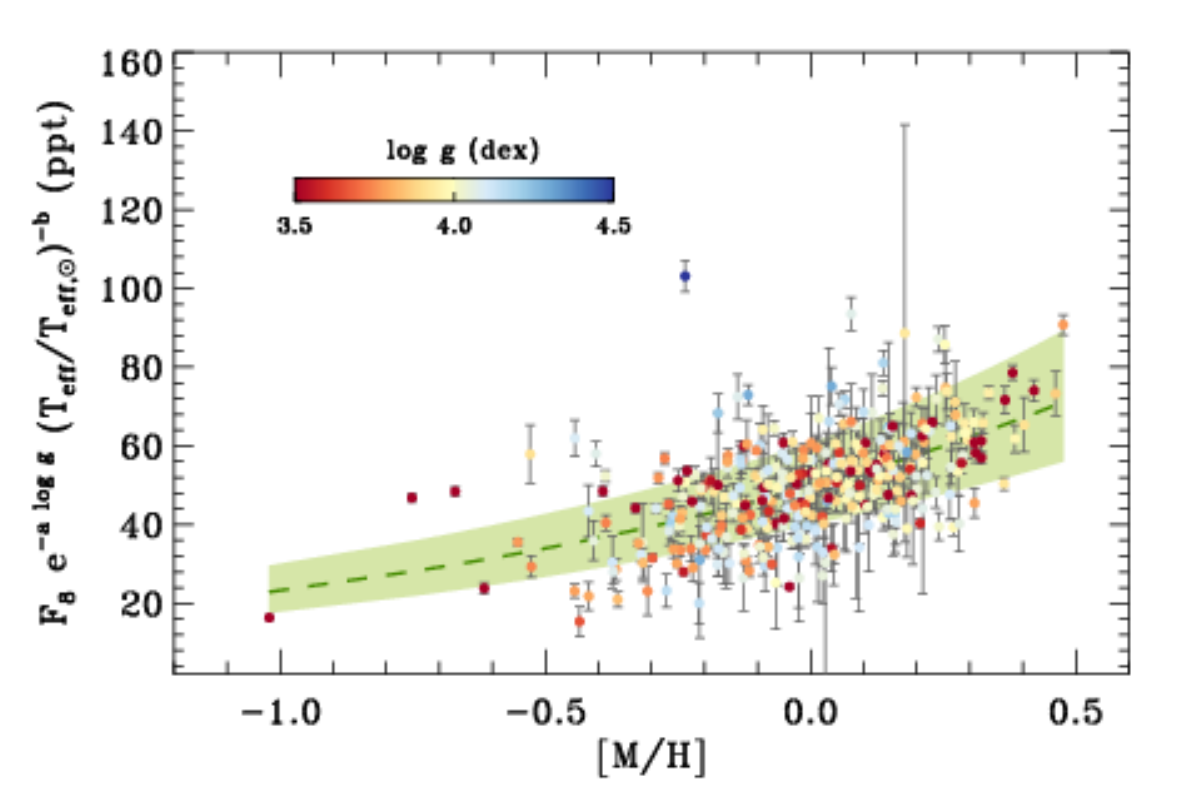}

\caption{The results of our fit of the dependence of flicker on spectroscopic observables. Similar to previous figures, we show plots focused on the 
surface gravity, temperature, and metallicity dependence (from top to bottom) for both the LC red giant stars (left panels) and the dwarfs and subgiants (right panels). Bayesian credible intervals of 68.3\,\% are shown with green shaded regions.}
\label{Fig:newSpec}
\end{figure}

The resulting fit for the dwarf and giant samples are obtained following the same approach used for Eq.~(\ref{eq:F8_astero}) and described in Sect.~\ref{sec:Bayes}. The results are shown in Figure \ref{Fig:newSpec}, and the exponents and their uncertainties are listed in Table \ref{tab:resultsSpec}. {It is worth mentioning that the differences in the fitted parameters between SC and LC sample are even more prononced when using the spectroscopic quantities, in particular with an inverted trend in the exponents related to temperature and metallicity. While in the case of the SC sample the trend with metallicity is in agreement with that found by C17 and our new fit using the asteroseismic quantities, for the LC sample this trend has an opposite direction, although the exponent is only slightly different than zero. We motivate this change as the combined effect of an offset $\ln \alpha$ having an opposite sign from SC to LC sample, a strong correlation between the exponent $b$ and $\ln \alpha$ (about 0.8), and the correlation between $b$ and the exponent $c$ of the metallicity term (corresponding to 0.46).}

The rms scatter of the spectroscopic fit was compared to the results from the seismic fit. This indicates that using only spectroscopic observables allows us to predict the flicker value to {6.6\,\% for the LC sample and 19.7\,\% for the SC sample, with an error likely dominated by the 0.02 dex uncertainty in surface gravity. This results in an overall precision arising from the fit to the spectroscopic relation with respect to the asteroseismic relation that is about 38\,\% and 47\,\% lower for the LC and SC samples, respectively.}


\begin{table}
\centering
 \caption{Median values of the inferred parameters $(a, b, c, \ln \alpha)$ for the spectroscopic fit for both SC {(343 stars)} and LC data {(2122 stars)}, with the physical parameter they relate to also indicated in brackets. Bayesian credible intervals of 68.3\,\% are included. 
 }
\begin{tabular}{lrrrr}
\hline\hline
\\[-8pt]
Model & \multicolumn{1}{c}{$a\,(log(g))$} & \multicolumn{1}{c}{$b\,(T_\mathrm{eff})$} & \multicolumn{1}{c}{$c$ ([M/H])} & \multicolumn{1}{c}{$\ln \alpha$}\\[1pt]
 \hline
 \\[-8pt]       
  $\mathrm{SC}$ & $-1.733^{+0.054}_{-0.050}$ & $2.321^{+0.212}_{-0.241}$ & $0.757^{+0.0510}_{-0.0530}$ & $3.904^{+0.203}_{-0.214}$\\ [1pt]
  $\mathrm{LC}$ & $-0.624^{+0.012}_{-0.014}$ & $-5.056^{+0.092}_{-0.038}$ & $-0.132^{+0.010}_{-0.012}$ &$-0.557^{+0.034}_{-0.066}$\\ [1pt]
  \hline
 \end{tabular}
\label{tab:resultsSpec}
\end{table}







\subsection{Bayesian Inference Approach}
\label{sec:Bayes}

For estimating the free parameters of the scaling laws given by Eq.~(\ref{eq:F8_astero}) and Eq.~(\ref{eq:F8_spec}), we adopt a Bayesian approach similar to that originally used by \cite{Corsaro:2013}, and subsequently by \cite{Bonanno2014} and by C17.

Considering Eq.~(\ref{eq:F8_spec}) as a reference, the actual fitting model that we need to consider is that obtained by converting the scaling relation to its natural logarithm, yielding
\begin{equation}
    \ln F_8^{\mathrm{(spec)}} = \ln \alpha + a \, \log g + b \ln\left(\frac{T_\mathrm{eff}}{T_\mathrm{eff,\odot}}\right) + c \mathrm{[M/H]} \, .
\end{equation}
{In this way, as shown by \cite{Corsaro:2013}, we can take into account the uncertainties on all the observed quantities, including the dependent variable $F_8$,} by implementing an analytical, parameter-dependent relation for the total relative uncertainty on the flicker predictions, which is given as
\begin{equation}
    \widetilde{\sigma}_{\mathrm{tot}}^2 \left(a,b,c\right) = \widetilde{\sigma}^2_{F_8} + a^2 \widetilde{\sigma}^2_\mathrm{\log g} + b^2 \widetilde{\sigma}^2_{T_\mathrm{eff}} + c^2 \widetilde{\sigma}^2_{\mathrm{[M/H]}}
\end{equation}
and includes the relative uncertainties on surface gravity, temperature and metallicity, on top of those of the measured flicker of stars. {Similarly to what reported by C17, we note that in this case $\widetilde{\sigma}_\mathrm{\log g}$ and $\widetilde{\sigma}_{\mathrm{[M/H]}}$ correspond to the formal uncertainties in $\log g$ and [M/H], respectively, since the two terms are logarithmic quantities used in the original scaling relation, and their uncertainties are therefore already in relative units.}

For the statistical inference on the flicker data we apply uniform priors on each free parameter of the scaling relations and consider a modified Gaussian log-likelihood of the form
\begin{equation}
\Lambda (\boldsymbol{\xi}) = \Lambda_0 (\boldsymbol{\xi}) - \frac{1}{2} \sum^N_{i=1} \left[ \frac{\Delta_i (\boldsymbol{\theta})}{\widetilde{\sigma}_\mathrm{tot,i} (\boldsymbol{\theta})} \right]^2 \, ,
\label{eq:likelihood}
\end{equation}
with $N$ the total number of flicker measurements, $\boldsymbol{\xi}$ the parameter vector of the free parameters (e.g. $(a,b,c)$, and $\Lambda_0 (\boldsymbol{\xi})$ an additional term depending only on the relative uncertainties, that is
\begin{equation}
\Lambda_0 (\boldsymbol{\xi}) = - \sum^N_{i=1} \ln \sqrt{2 \pi} \widetilde{\sigma}_\mathrm{tot,i} (\boldsymbol{\xi}) \, .
\end{equation}
Finally, as shown in C17, we note that the residuals between observed and predicted flicker values are defined as
\begin{equation}
\Delta_i (\boldsymbol{\xi}) = \ln F_8^\mathrm{obs} - \ln F_8^\mathrm{pred} (\boldsymbol{\xi}) \, .
\end{equation}

A perfectly analogous approach is used for the scaling law given by Eq.~(\ref{eq:F8_astero}), in which the total relative uncertainty obtained is given as
\begin{equation}
    \widetilde{\sigma}_{\mathrm{tot}}^2 \left(s,t,u\right) = \widetilde{\sigma}^2_{F_8} + s^2 \widetilde{\sigma}^2_{\nu_\mathrm{max}} + t^2 \widetilde{\sigma}^2_{M} + u^2 \widetilde{\sigma}^2_{\mathrm{[M/H]}} \, .
\end{equation}



\subsection{Application to the full APOGEE DR14 sample}
\begin{figure}[!ht]
    \centering
    \includegraphics[width=0.495\linewidth]{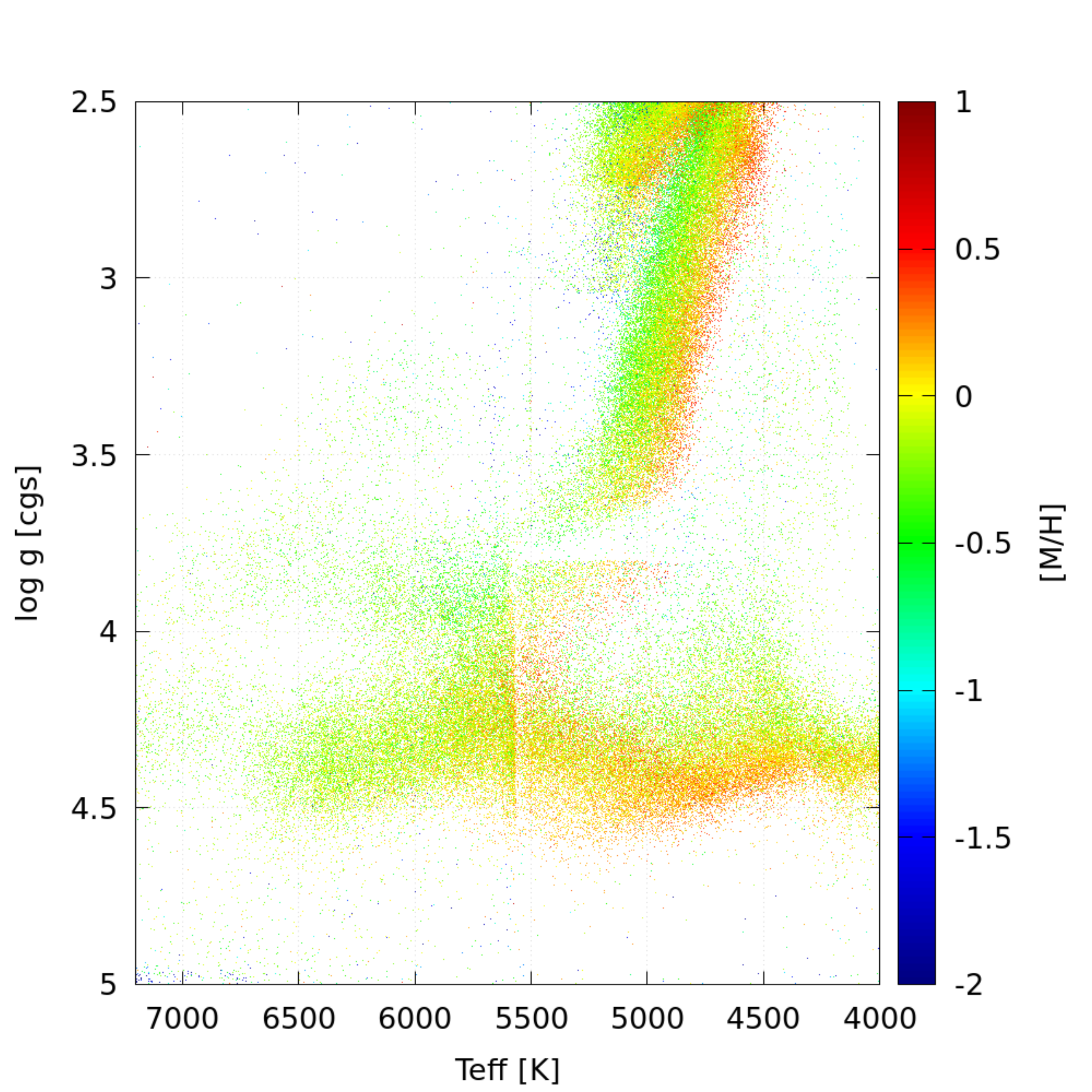}
    \includegraphics[width=0.495\linewidth]{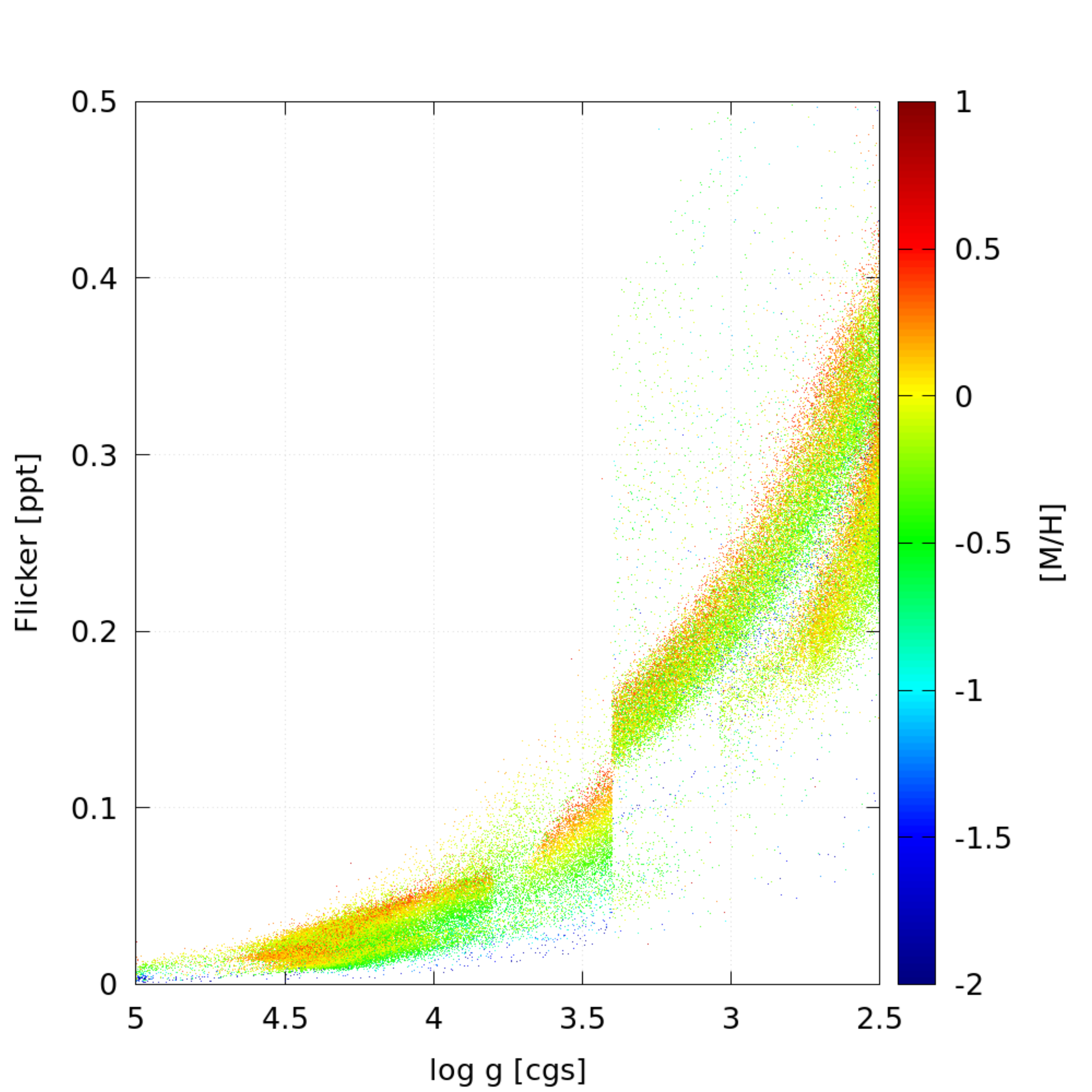}
    \caption{(a) HR diagram of stars in our final APOGEE sample. (b) $F_8$ values inferred from $\log g$ and [M/H] (dependence on $T_{\rm eff}$ not explicitly shown) using our empirical relations. Note the discontinuity in the $F_8$ relation (panel b) at $\log g = 3.4$ is due to the use of our two different flicker relations above and below that value. Note also the separation of red clump stars from red giant stars in both panels.}
    \label{fig:full_hrd}
\end{figure}

Using our new empirical relations between $F_8$ and spectroscopic stellar parameters, in Table~\ref{tab:apf8} we report $F_8$ for all 129,055 APOGEE DR14 stars having $T_{\rm eff}$, $\log g$, and [M/H] measurements, although we caution that flicker is only formally valid for 4500~K $< T_{\rm eff} <$ 7150~K and $2.5 < \log g  < 4.6$, and in fact the flicker-gravity relation turns over at $\log g < 2.5$ as the timescale of granulation shifts out of the frequency window measured by flicker.
A Hertzsprung-Russell diagram for these stars is presented in Figure~\ref{fig:full_hrd}(a), and the resulting $F_8$ as a function of $\log g$ and [M/H] is shown in Figure~\ref{fig:full_hrd}(b).

\begin{table}[tbp]
\footnotesize
\caption{Results of predicted Flicker ($F_8$) and RV jitter for the full APOGEE sample. The full table is provided in the electronic version of the journal. A portion is shown here for guidance regarding form and content.
\textbf{a} {\footnotesize Stellar parameters from APOGEE DR14 catalog.}
\textbf{b} {\footnotesize $F_8$ estimated from stellar spectroscopic parameters using new empirical relation (Sec.~\ref{sec:results}).}
\textbf{d} {\footnotesize RV jitter amplitude estimated from empirical Flicker-jitter relation of \citet{Bastien:2014}. Uncertainty includes error contributions from both $F_8$ and the Flicker-jitter relation.}
\textbf{e} {\footnotesize RV jitter amplitude estimated from empirical Flicker-jitter relation of \citet{Oshagh:2017}. Uncertainty includes error contributions from both $F_8$ and the Flicker-jitter relation.}
\textbf{f} {\footnotesize RV jitter amplitude estimated from empirical Flicker-jitter relation of \citet{Cegla:2014}. Uncertainty includes error contributions from both $F_8$ and the Flicker-jitter relation.}
\textbf{g} {\footnotesize Light curve variability amplitude from KELT \citep{Oelkers:2018}. RV jitter estimates for stars with variability amplitudes $\gtrsim$10~mmag should be treated as lower limits; see the text.}}

\begin{center}
\begin{tabular}{cccccccccc}
\tableline
2MASS ID & $T_{\rm eff}$\tablenotemark{a} & $\log g$\tablenotemark{a} & [M/H]\tablenotemark{a} & $F_8$\tablenotemark{b} & [C/N]\tablenotemark{a} & 
RV$_1$\tablenotemark{d} & RV$_2$\tablenotemark{e} & RV$_3$\tablenotemark{f} & LC r.m.s.\tablenotemark{g} \\
& (K) & (cgs) &  & (ppt) &  
& (m~s$^{-1}$) & (m~s$^{-1}$) & (m~s$^{-1}$) & (mag) \\
\tableline
00001977$-$2003393 & 5351 & 4.14 & $-$0.223 & 0.026 & \phm{$-$}0.004 & 4.31 $\pm$ 1.69 & 5.71 $\pm$ 3.02 & 1.46 $\pm$ 0.54 & \nodata \\
00002021$+$6302567 & 4933 & 2.52 & $-$0.017 & 0.262 & $-$0.586 & 11.85 $\pm$ 1.53 & 10.30 $\pm$ 3.90 & 5.71 $\pm$ 0.21 & 0.01 \\
00002035$+$6250406 & 5049 & 2.56 & \phm{$-$}0.217 & 0.220 & $-$0.098 & 10.50 $\pm$ 1.38 & 9.47 $\pm$ 3.59 & 4.95 $\pm$ 0.22 & \nodata \\
00002038$-$1912052 & 6172 & 4.41 & $-$0.010 & 0.027 & \phm{$-$}0.007 & 4.33 $\pm$ 1.67 & 5.72 $\pm$ 3.00 & 1.47 $\pm$ 0.54 & \nodata \\
00002141$+$8606336 & 4856 & 3.19 & \phm{$-$}0.094 & 0.184 & $-$0.213 & 9.37 $\pm$ 1.26 & 8.79 $\pm$ 3.34 & 4.31 $\pm$ 0.23 & \nodata \\
00002142$-$1929009 & 5357 & 4.54 & \phm{$-$}0.069 & 0.016 & \phm{$-$}0.008 & 3.99 $\pm$ 2.45 & 5.51 $\pm$ 3.91 & 1.27 $\pm$ 0.77 & \nodata \\
00002338$+$6141442 & 4987 & 2.52 & $-$0.061 & 0.249 & $-$0.445 & 11.43 $\pm$ 1.48 & 10.04 $\pm$ 3.80 & 5.47 $\pm$ 0.21 & \nodata \\
00002388$+$6151472 & 4785 & 3.01 & \phm{$-$}0.199 & 0.219 & $-$0.391 & 10.47 $\pm$ 1.37 & 9.46 $\pm$ 3.58 & 4.93 $\pm$ 0.22 & 0.017 \\
00002561$+$5528511 & 5512 & 3.82 & \phm{$-$}0.010 & 0.059 & \phm{$-$}0.003 & 5.37 $\pm$ 1.11 & 6.35 $\pm$ 2.61 & 2.05 $\pm$ 0.34 & \nodata \\
00002596$+$7433383 & 4952 & 3.09 & $-$0.114 & 0.182 & $-$0.276 & 9.30 $\pm$ 1.25 & 8.75 $\pm$ 3.33 & 4.27 $\pm$ 0.23 & \nodata \\


\tableline

\end{tabular}
\end{center}

\label{tab:apf8}
\end{table}

Table~\ref{tab:apf8} also provides the [C/N] values from the APOGEE DR14 catalog so that we can identify stars in our sample that may be Red Clump (RC) stars. While the granulation properties of RC stars may differ slightly from those of their first-ascent red giant cousins, the general relationships between $F_8$ and other stellar parameters should still hold \citep[see, e.g.,][]{Bastien:2016}, and we expect that the RV jitter that we predict from $F_8$ using our empirical relationships should hold as well. Stars in Table~\ref{tab:apf8} can be flagged as likely RC if they meet the following criteria, as defined by \citet{Holtzman:2018}: 
$2.38 < \log g < 3.5$, and [C/N] $> -0.08 -0.5$[M/H] $-0.0039\Delta T$, where $\Delta T \equiv T_{\rm eff}-\left\{4444.14 + 554.311(\log g - 2.5) - 307.962 \rm{[M/H]} \right\}$. In Figure \ref{fig:full_hrd}b, these core helium burning stars, specifically the slightly more massive secondary clump stars \citep{Girardi1999}, are responsible for the lower branch of flicker values that appears below a $\log g$ of about 3.

\subsection{Predicting RV Jitter from Flicker}



To date, there have not been large datasets published that provide both observed measures of $F_8$ and of RV jitter from which empirical \flicker-jitter relations can be robustly established. However, using smaller samples, three studies have attempted to provide a correlation between $F_8$ and jitter, so we use all three to estimate RV jitter for our sample in Table~\ref{tab:apf8}.

The first is from \citet{Bastien:2014}, based on a sample of relatively quiet stars having overall photometric variability of $\lesssim$3~ppt: 
$RV_{\rm rms} = (31.99\pm3.95) \times F_8/{\rm ppt} + (3.46\pm1.19)$ m~s$^{-1}$. 
The second is from \citet{Oshagh:2017}, which gives a similar (but less precise) fit, but also extends to stars with slightly larger amplitude photometric variability ($\lesssim$10~ppt): 
$RV_{\rm rms} = (19.49\pm7.34) \times F_8/{\rm ppt} + (5.19\pm2.12)$ m~s$^{-1}$. 
Finally, the third relation is from \citet{Cegla:2014}, 
which actually used chromospheric activity as a proxy to estimate the RV jitter, but still provides a comparable relation: 
$RV_{\rm rms} = 18.04 \times F_8/{\rm ppt} + 0.98$ m~s$^{-1}$. 

\begin{figure}[!ht]
    \centering
    \includegraphics[width=0.5\linewidth]{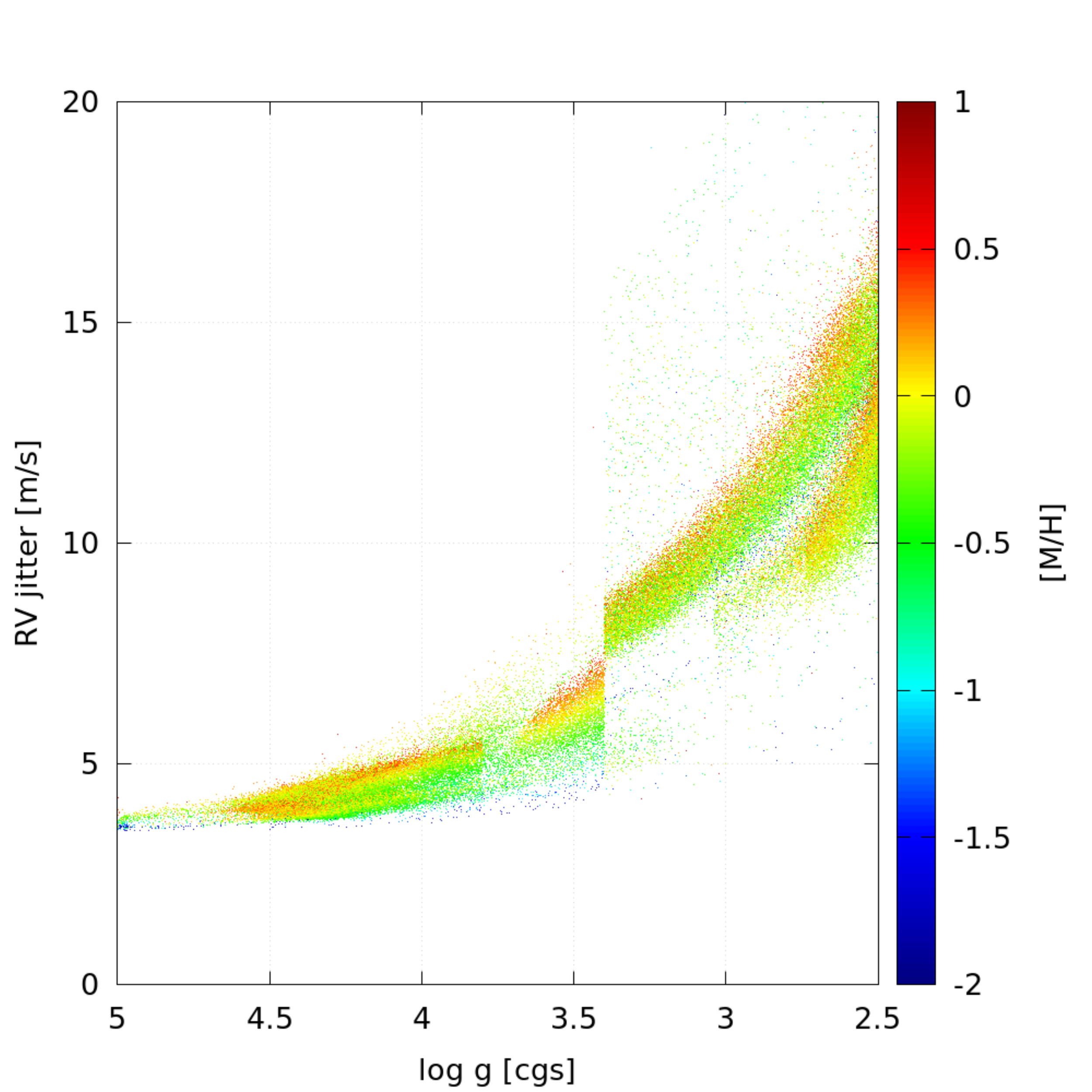}
    \caption{RV jitter estimated from $F_8$ using the empirical Flicker-jitter relation of \citet{Bastien:2014}. Note that Table~\ref{tab:apf8} also reports estimated RV jitter values using the similar relations from \citet{Oshagh:2017} and from \citet{Cegla:2014}.}
    \label{fig:final_rv}
\end{figure}

All three relations yield comparable estimates for the RV jitter. In Fig.~\ref{fig:final_rv} we present the resulting RV jitter estimates specifically using the \citet{Bastien:2014} relation.

\section{Discussion}\label{sec:discussion}

\subsection{Breakdown of predicted RV jitter based on activity}
For stars similar to those in our study, 
previous work has found that plage and/or spot driven variations begin to dominate the RV jitter when their photometric variations exceed $\sim$10~ppt \citep{Oshagh:2017}, which is well within reach of most all-sky surveys of photometric variability \citep[e.g.,][]{Oelkers:2018}. We flag these active stars in Table \ref{tab:apf8} based on the light curve variability reported by \citet{Oelkers:2018} using the KELT \citep{Pepper:2007, Pepper:2012} survey, as these stars will likely display radial velocity jitter significantly above the predictions based on their granulation. However, other tools have been developed to estimate the plage/spot-driven RV jitter \citep[e.g.,][]{Aigrain:2012,Dumusque:2014}. Therefore, it should be possible now to reliably estimate RV jitter for stars over very large swathes of the HR diagram (see Fig.~\ref{fig:full_hrd}a) and with photometric variability amplitudes from the granulation-dominated regime (0.01--10~ppt) to the plage- and spot-dominated regime ($\gtrsim$10~ppt). In particular, it should be possible to identify stars whose expected RV jitter is reliably below a few m~s$^{-1}$ (Fig.~\ref{fig:final_rv}).

\subsection{General guidelines and caveats} 

The ability to accurately infer the amplitude of granulation ``flicker", including the dependence on metallicity, from spectroscopic parameters is important because upcoming all-sky spectroscopic surveys will permit characterization of the granulation noise for large numbers of stars that may be inaccessible for direct measurement of $F_8$ by, e.g., {\it TESS}. 
In addition, traditional methods for determining the activity of stars, which may reach sensitivity to photometric variations as low as a few ppt, will not in general be sensitive to the low-amplitude granulation-driven variations, which only reach photometric amplitudes of $\lesssim$0.5~ppt (see Fig.~\ref{fig:full_hrd}b).

Indeed, past attempts to screen out stars likely to exhibit high RV jitter have found that some otherwise ``quiet" stars still exhibit RV jitter of up to $\sim$20~m~s$^{-1}$ \citep[e.g.,][]{Wright:2005,Isaacson:2010}. This can be explained as arising from granulation-driven RV jitter \citep[e.g.,][]{Bastien:2014,Cegla:2014,Oshagh:2017}, which can reach up to $\sim$20~m~s$^{-1}$ (Fig.~\ref{fig:final_rv}), despite the granulation-driven photometric variations reaching only up to $\sim$0.5~ppt (Fig.~\ref{fig:full_hrd}b).

With the relationships discussed here, it should now be possible to estimate the predicted light curve flicker and radial velocity jitter from a single spectrum, without committing significant observing resources to a target. However, there are regions of parameter space where these relationships should not be applied without caution. The first is stars with significant activity, as discussed above. In that case, our predictions are likely to be lower limits on the actual amount of jitter. 

The second group of stars where these predictions should be treated with caution are those stars significantly outside our calibration sample. M dwarfs, in particular, have inaccurate surface gravities from APOGEE, higher surface gravities than the stars in our sample, and are outside of the regime where where flicker is considered a reliable tracer of surface gravity \citep{Bastien:2016}, and therefore our estimations of jitter, which are based on flicker, are unlikely to be accurate for these stars. Similarly, there are no hot main sequence stars in our calibration sample, and we expect these stars with shallow or nonexistant surface convection zones to have very different radial velocity variability than the stars studied here. Finally, the flicker technique becomes double valued for stars with a surface gravity below 2.5 dex, as the timescales of granulation and oscillations shift relative to the frequency window where flicker is calibrated. Because of this, we advise caution when applying our jitter relation to low gravity giants, as work by \citet{Hekker2008} suggests that our predictions of jitter do not grow steeply enough with surface gravity for these stars. 


\subsection{Alternatives to Flicker}
Because of the importance of predicting the stellar variability and radial velocity jitter, a number of authors have explored relationships to predict these variables. We discuss a few recent studies and compare them to our analysis.

\subsubsection{\citet{Pande2018}}
The methodology presented by \cite{Pande2018} is based on the tight relation between \numax\,and the stellar surface gravity via the granulation-driven signal. In particular, the authors show how to quickly estimate $\log g$ for many stars, with a relative good precision, by measuring the granulation power from the stellar power spectra starting from the knowledge of \numax\ for oscillating stars. However, they only take into account the dependency of granulation on $T_\mathrm{eff}$ and not on metallicity. In addition, the granulation power is corrected by the noise level in the power spectra, which is not estimated for each star but extrapolated from a calibration to a benchmark sample, a step that is not required when measuring the flicker amplitude directly. As a result, the estimates of $\log g$ obtained by \cite{Pande2018}, will not be as accurate as one could obtain when including the metallicity dependence of stellar granulation (see also \citealt{Stassun:2018}) from the flicker measurement, as we show here. Finally, we note that the methodology presented by \cite{Pande2018} can only be applied to stars showing oscillations, while a flicker-based estimate of $\log g$ is not subject to this limitation.

\subsubsection{\citet{Ness2018}}
In \citet{Ness2018} a polynomial fit is made to the autocorrelation function of \kepler\ giants, and the value at each frequency for stars of known properties is used to determine the sensitivity of that frequency to the observables including temperature, surface gravity, and metallicity. In that analysis, the authors find no significant dependence of the autocorrelation at any frequency, including those used to calculate flicker, to the stellar metallicity. This is in conflict with our results, which find a small but significant dependence of the flicker amplitude to convection for giant stars. We note, however, that the \citet{Ness2018} analysis used stars covering a wider range in surface gravity, including low values where the dependence of flicker on surface gravity reverses, as well as stars on both the clump and giant branch, where the dependence of flicker on surface gravity is different \citep{Bastien:2016}. We speculate that these two effects 
could conspire to hide the relatively weak metallicity dependence we measure here. It would be interesting to see whether the \citet{Ness2018} analysis would provide a metallicity dependence more consistent with our results if done over the restricted range of temperature, surface gravity, and evolutionary state used here.



\subsubsection{\citet{Yu:2018}}
\begin{figure}[!ht]
    \centering
    \includegraphics[width=0.5\linewidth,clip,trim=0 0 0 75pt]{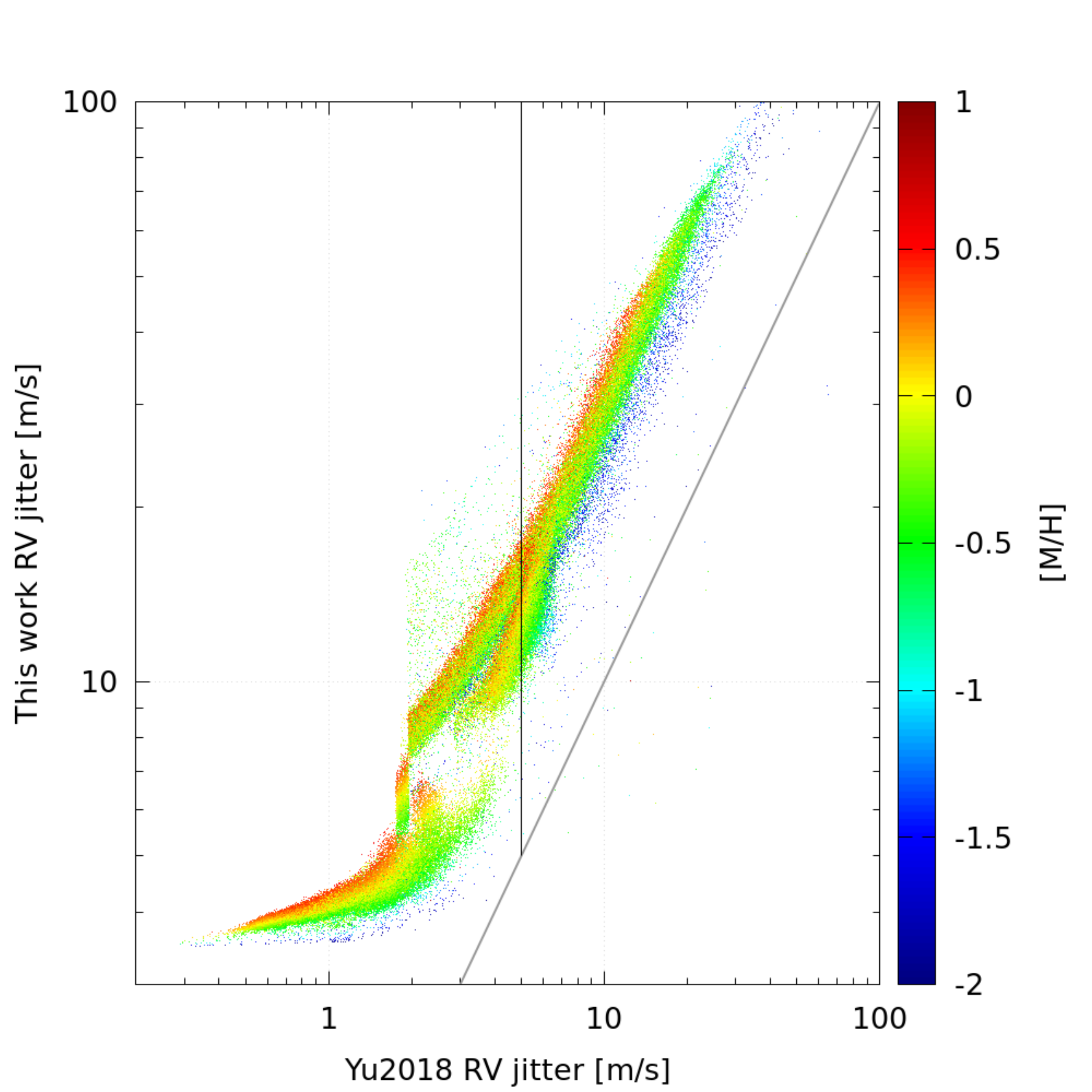}
    \caption{Comparison of RV jitter predicted in this work from granulation versus that predicted by \citet{Yu:2018} from stellar oscillations. A line representing 1-to-1 agreement is shown, as well as a vertical line corresponding approximately to the range for which the \flicker\ relation is formally valid ($\log g > 2.5$, corresponding to RV jitter below $\sim$5~m~s$^{-1}$ on the horizontal axis). Discrepancies in the predictions for RGB and clump stars can be seen around \citet{Yu:2018} values of 4~m~s$^{-1}$, and the break between our dwarf and giant predictions is visible around our values of 7~m~s$^{-1}$.}
    \label{fig:jitter_comp}
\end{figure}

Recently, \citet{Yu:2018} have developed an approach to RV jitter arising from stellar oscillations, and they provide functions that enable the oscillation-driven jitter to be estimated from observables such as $T_{\rm eff}$ and $\log g$, though they do not include a metallicity dependence. They also note that their estimates are lower limits to the RV jitter because they specifically do not include the effects of granulation flicker, as we have done here. Nonetheless, it is instructive to compare the RV jitter predicted here from granulation flicker versus that predicted by \citet{Yu:2018} from oscillations. The comparison is shown in Figure~\ref{fig:jitter_comp}, where we have used Equation~7 from \citet{Yu:2018}. As expected, the granulation-driven jitter is in general larger than that predicted from oscillations alone, and moreover there is a clear metallicity dependence arising from our explicit inclusion of metallicity as a term in the predicted granulation flicker amplitudes.

\section{Summary and Conclusions}\label{sec:summary}
As we push to identify and characterize smaller and smaller planets, the existence of stellar noise becomes more and more significant, impacting our ability to measure both the light curve and the radial velocity variation. Here, we update the relationships between stellar properties and convective granulation discussed in \citet{Corsaro:2017}, and show that the metallicity dependence is somewhat weaker than what was observed in that work. We also show that both the light-curve granulation flicker and its associated radial-velocity jitter can be predicted to better than 7 percent for giants and 20 percent for dwarfs and subgiants without asteroseismology, using only spectroscopic observables like effective temperature, metallicity, and surface gravity. 
We have applied our new relations to 129,000 stars observed by the SDSS APOGEE survey, and report for them the spectroscopically estimated flicker and jitter. Finally, we identify stars whose total jitter is likely to be even larger by virtue of large-amplitude photometric variability. 

These spectroscopic estimates of flicker and jitter can be used to improve estimates of planet frequency as a function of host star type, as the radial velocity jitter changes the size and location of planets that can be detected as a function of stellar mass and metallicity.  Additionally, these predictions will allow estimation of future planet detectability before committing significant follow-up resources, as detecting planets below the level of radial velocity jitter requires significant numbers of carefully planned observations. Future work that predicts the timescale as well as the amplitude of the radial velocity jitter as a function of stellar properties would also enhance the ability to detect small planets, and we therefore encourage efforts to efficiently identify the timescales of stellar granulation \citep[e.g.][]{Kallinger2016,Bugnet2018} and the publication of larger datasets of RV jitter values for stars of all types.

\acknowledgements
We thank S. Mathur, R. Garc{\'i}a, M. Pinsonneault, J. Yu,  D. Stello,  S. Hekker, R. Siverd, D. Huber, and A. Chontos for helpful discussions.

Support for this work was provided by NASA through the NASA Hubble Fellowship grant \#51424 awarded by the Space Telescope Science Institute, which is operated by the Association of Universities for Research in Astronomy, Inc., for NASA, under contract NAS5-26555. 
K.G.S.\ acknowledges NASA grant 17-XRP17 2-0024. 
This work made use of the Filtergraph data visualization service \citep{Burger:2013}, supported by the Vanderbilt Center for Autism \& Innovation.
E.C.\ is funded by the European Union’s Horizon 2020 research and innovation program under the Marie Sklodowska-Curie grant agreement No.~664931.

Funding for the Sloan Digital Sky Survey IV has been provided by the Alfred P.\ Sloan Foundation, the U.S.\ Department of Energy Office of Science, and the Participating Institutions. SDSS-IV acknowledges
support and resources from the Center for High-Performance Computing at
the University of Utah. The SDSS web site is www.sdss.org.
SDSS-IV is managed by the Astrophysical Research Consortium for the 
Participating Institutions of the SDSS Collaboration including the 
Brazilian Participation Group, the Carnegie Institution for Science, 
Carnegie Mellon University, the Chilean Participation Group, the French Participation Group, Harvard-Smithsonian Center for Astrophysics, 
Instituto de Astrof\'isica de Canarias, The Johns Hopkins University, 
Kavli Institute for the Physics and Mathematics of the Universe (IPMU) / 
University of Tokyo, the Korean Participation Group, Lawrence Berkeley National Laboratory, 
Leibniz Institut f\"ur Astrophysik Potsdam (AIP),  
Max-Planck-Institut f\"ur Astronomie (MPIA Heidelberg), 
Max-Planck-Institut f\"ur Astrophysik (MPA Garching), 
Max-Planck-Institut f\"ur Extraterrestrische Physik (MPE), 
National Astronomical Observatories of China, New Mexico State University, 
New York University, University of Notre Dame, 
Observat\'ario Nacional / MCTI, The Ohio State University, 
Pennsylvania State University, Shanghai Astronomical Observatory, 
United Kingdom Participation Group,
Universidad Nacional Aut\'onoma de M\'exico, University of Arizona, 
University of Colorado Boulder, University of Oxford, University of Portsmouth, 
University of Utah, University of Virginia, University of Washington, University of Wisconsin, 
Vanderbilt University, and Yale University.

\bibliographystyle{aasjournal} 
\bibliography{ms}

\begin{thebibliography}{}
\expandafter\ifx\csname natexlab\endcsname\relax\def\natexlab#1{#1}\fi
\providecommand{\url}[1]{\href{#1}{#1}}

\bibitem[{{Abolfathi} {et~al.}(2017){Abolfathi}, {Aguado}, {Aguilar}, {Allende
  Prieto}, {Almeida}, {Tasnim Ananna}, {Anders}, {Anderson}, {Andrews},
  {Anguiano}, \& et~al.}]{DR14}
{Abolfathi}, B., {Aguado}, D.~S., {Aguilar}, G., {et~al.} 2017, ArXiv e-prints,
  arXiv:1707.09322

\bibitem[{{Aerts} {et~al.}(2010){Aerts}, {Christensen-Dalsgaard}, \&
  {Kurtz}}]{Aerts2010}
{Aerts}, C., {Christensen-Dalsgaard}, J., \& {Kurtz}, D.~W. 2010,
  {Asteroseismology}

\bibitem[{{Aigrain} {et~al.}(2012){Aigrain}, {Pont}, \&
  {Zucker}}]{Aigrain:2012}
{Aigrain}, S., {Pont}, F., \& {Zucker}, S. 2012, \mnras, 419, 3147

\bibitem[{{Auvergne} {et~al.}(2009){Auvergne}, {Bodin}, {Boisnard}, {Buey},
  {Chaintreuil}, {Epstein}, {Jouret}, {Lam-Trong}, {Levacher}, {Magnan},
  {Perez}, {Plasson}, {Plesseria}, {Peter}, {Steller}, {Tiph{\`e}ne}, {Baglin},
  {Agogu{\'e}}, {Appourchaux}, {Barbet}, {Beaufort}, {Bellenger}, {Berlin},
  {Bernardi}, {Blouin}, {Boumier}, {Bonneau}, {Briet}, {Butler}, {Cautain},
  {Chiavassa}, {Costes}, {Cuvilho}, {Cunha-Parro}, {de Oliveira Fialho},
  {Decaudin}, {Defise}, {Djalal}, {Docclo}, {Drummond}, {Dupuis}, {Exil},
  {Faur{\'e}}, {Gaboriaud}, {Gamet}, {Gavalda}, {Grolleau}, {Gueguen},
  {Guivarc'h}, {Guterman}, {Hasiba}, {Huntzinger}, {Hustaix}, {Imbert},
  {Jeanville}, {Johlander}, {Jorda}, {Journoud}, {Karioty}, {Kerjean},
  {Lafond}, {Lapeyrere}, {Landiech}, {Larqu{\'e}}, {Laudet}, {Le Merrer},
  {Leporati}, {Leruyet}, {Levieuge}, {Llebaria}, {Martin}, {Mazy}, {Mesnager},
  {Michel}, {Moalic}, {Monjoin}, {Naudet}, {Neukirchner}, {Nguyen-Kim},
  {Ollivier}, {Orcesi}, {Ottacher}, {Oulali}, {Parisot}, {Perruchot},
  {Piacentino}, {Pinheiro da Silva}, {Platzer}, {Pontet}, {Pradines},
  {Quentin}, {Rohbeck}, {Rolland}, {Rollenhagen}, {Romagnan}, {Russ}, {Samadi},
  {Schmidt}, {Schwartz}, {Sebbag}, {Smit}, {Sunter}, {Tello}, {Toulouse},
  {Ulmer}, {Vandermarcq}, {Vergnault}, {Wallner}, {Waultier}, \&
  {Zanatta}}]{Auvergne:2009}
{Auvergne}, M., {Bodin}, P., {Boisnard}, L., {et~al.} 2009, \aap, 506, 411

\bibitem[{{Bastien} {et~al.}(2013){Bastien}, {Stassun}, {Basri}, \&
  {Pepper}}]{Bastien:2013}
{Bastien}, F.~A., {Stassun}, K.~G., {Basri}, G., \& {Pepper}, J. 2013, \nat,
  500, 427

\bibitem[{{Bastien} {et~al.}(2016){Bastien}, {Stassun}, {Basri}, \&
  {Pepper}}]{Bastien:2016}
---. 2016, \apj, 818, 43

\bibitem[{{Bastien} {et~al.}(2014){Bastien}, {Stassun}, {Pepper}, {Wright},
  {Aigrain}, {Basri}, {Johnson}, {Howard}, \& {Walkowicz}}]{Bastien:2014}
{Bastien}, F.~A., {Stassun}, K.~G., {Pepper}, J., {et~al.} 2014, \aj, 147, 29

\bibitem[{{Beatty} \& {Gaudi}(2015)}]{BeattyGaudi2015}
{Beatty}, T.~G., \& {Gaudi}, B.~S. 2015, \pasp, 127, 1240

\bibitem[{{Berger} {et~al.}(2018){Berger}, {Huber}, {Gaidos}, \& {van
  Saders}}]{Berger:2018}
{Berger}, T.~A., {Huber}, D., {Gaidos}, E., \& {van Saders}, J.~L. 2018, ArXiv
  e-prints, arXiv:1805.00231

\bibitem[{{Blanton} {et~al.}(2017){Blanton}, {Bershady}, {Abolfathi},
  {Albareti}, {Allende Prieto}, {Almeida}, {Alonso-Garc{\'{\i}}a}, {Anders},
  {Anderson}, {Andrews}, \& et~al.}]{Blanton2017}
{Blanton}, M.~R., {Bershady}, M.~A., {Abolfathi}, B., {et~al.} 2017, \aj, 154,
  28

\bibitem[{{Bonanno} {et~al.}(2014){Bonanno}, {Corsaro}, \&
  {Karoff}}]{Bonanno2014}
{Bonanno}, A., {Corsaro}, E., \& {Karoff}, C. 2014, A\&A, 571, A35

\bibitem[{{Borucki} {et~al.}(2010){Borucki}, {Koch}, {Basri}, {Batalha},
  {Brown}, {Caldwell}, {Caldwell}, {Christensen-Dalsgaard}, {Cochran},
  {DeVore}, {Dunham}, {Dupree}, {Gautier}, {Geary}, {Gilliland}, {Gould},
  {Howell}, {Jenkins}, {Kondo}, {Latham}, {Marcy}, {Meibom}, {Kjeldsen},
  {Lissauer}, {Monet}, {Morrison}, {Sasselov}, {Tarter}, {Boss}, {Brownlee},
  {Owen}, {Buzasi}, {Charbonneau}, {Doyle}, {Fortney}, {Ford}, {Holman},
  {Seager}, {Steffen}, {Welsh}, {Rowe}, {Anderson}, {Buchhave}, {Ciardi},
  {Walkowicz}, {Sherry}, {Horch}, {Isaacson}, {Everett}, {Fischer}, {Torres},
  {Johnson}, {Endl}, {MacQueen}, {Bryson}, {Dotson}, {Haas}, {Kolodziejczak},
  {Van Cleve}, {Chandrasekaran}, {Twicken}, {Quintana}, {Clarke}, {Allen},
  {Li}, {Wu}, {Tenenbaum}, {Verner}, {Bruhweiler}, {Barnes}, \&
  {Prsa}}]{Borucki:2010}
{Borucki}, W.~J., {Koch}, D., {Basri}, G., {et~al.} 2010, Science, 327, 977

\bibitem[{{Brown} {et~al.}(1991){Brown}, {Gilliland}, {Noyes}, \&
  {Ramsey}}]{Brown1991}
{Brown}, T.~M., {Gilliland}, R.~L., {Noyes}, R.~W., \& {Ramsey}, L.~W. 1991,
  \apj, 368, 599

\bibitem[{{Bugnet} {et~al.}(2018){Bugnet}, {Garc{\'{\i}}a}, {Davies}, {Mathur},
  {Corsaro}, {Hall}, \& {Rendle}}]{Bugnet2018}
{Bugnet}, L., {Garc{\'{\i}}a}, R.~A., {Davies}, G.~R., {et~al.} 2018, \aap,
  620, A38

\bibitem[{{Burger} {et~al.}(2013){Burger}, {Stassun}, {Pepper}, {Siverd},
  {Paegert}, {De Lee}, \& {Robinson}}]{Burger:2013}
{Burger}, D., {Stassun}, K.~G., {Pepper}, J., {et~al.} 2013, Astronomy and
  Computing, 2, 40

\bibitem[{{Cegla} {et~al.}(2014){Cegla}, {Stassun}, {Watson}, {Bastien}, \&
  {Pepper}}]{Cegla:2014}
{Cegla}, H.~M., {Stassun}, K.~G., {Watson}, C.~A., {Bastien}, F.~A., \&
  {Pepper}, J. 2014, \apj, 780, 104

\bibitem[{{Chaplin} {et~al.}(2011){Chaplin}, {Kjeldsen},
  {Christensen-Dalsgaard}, {Basu}, {Miglio}, {Appourchaux}, {Bedding},
  {Elsworth}, {Garc{\'{\i}}a}, {Gilliland}, {Girardi}, {Houdek}, {Karoff},
  {Kawaler}, {Metcalfe}, {Molenda-{\.Z}akowicz}, {Monteiro}, {Thompson},
  {Verner}, {Ballot}, {Bonanno}, {Brand{\~a}o}, {Broomhall}, {Bruntt},
  {Campante}, {Corsaro}, {Creevey}, {Do{\u g}an}, {Esch}, {Gai}, {Gaulme},
  {Hale}, {Handberg}, {Hekker}, {Huber}, {Jim{\'e}nez}, {Mathur}, {Mazumdar},
  {Mosser}, {New}, {Pinsonneault}, {Pricopi}, {Quirion}, {R{\'e}gulo},
  {Salabert}, {Serenelli}, {Silva Aguirre}, {Sousa}, {Stello}, {Stevens},
  {Suran}, {Uytterhoeven}, {White}, {Borucki}, {Brown}, {Jenkins}, {Kinemuchi},
  {Van Cleve}, \& {Klaus}}]{Chaplin2011}
{Chaplin}, W.~J., {Kjeldsen}, H., {Christensen-Dalsgaard}, J., {et~al.} 2011,
  Science, 332, 213

\bibitem[{{Corsaro} {et~al.}(2013){Corsaro}, {Fr{\"o}hlich}, {Bonanno},
  {Huber}, {Bedding}, {Benomar}, {De Ridder}, \& {Stello}}]{Corsaro:2013}
{Corsaro}, E., {Fr{\"o}hlich}, H.-E., {Bonanno}, A., {et~al.} 2013, \mnras,
  430, 2313

\bibitem[{{Corsaro} {et~al.}(2017){Corsaro}, {Mathur}, {Garc{\'{\i}}a},
  {Gaulme}, {Pinsonneault}, {Stassun}, {Stello}, {Tayar}, {Trampedach},
  {Jiang}, {Nitschelm}, \& {Salabert}}]{Corsaro:2017}
{Corsaro}, E., {Mathur}, S., {Garc{\'{\i}}a}, R.~A., {et~al.} 2017, \aap, 605,
  A3

\bibitem[{{Dumusque} {et~al.}(2014){Dumusque}, {Boisse}, \&
  {Santos}}]{Dumusque:2014}
{Dumusque}, X., {Boisse}, I., \& {Santos}, N.~C. 2014, \apj, 796, 132

\bibitem[{{Dumusque} {et~al.}(2011){Dumusque}, {Santos}, {Udry}, {Lovis}, \&
  {Bonfils}}]{Dumusque:2011}
{Dumusque}, X., {Santos}, N.~C., {Udry}, S., {Lovis}, C., \& {Bonfils}, X.
  2011, \aap, 527, A82

\bibitem[{{Garc{\'{\i}}a P{\'e}rez} {et~al.}(2016){Garc{\'{\i}}a P{\'e}rez},
  {Allende Prieto}, {Holtzman}, {Shetrone}, {M{\'e}sz{\'a}ros}, {Bizyaev},
  {Carrera}, {Cunha}, {Garc{\'{\i}}a-Hern{\'a}ndez}, {Johnson}, {Majewski},
  {Nidever}, {Schiavon}, {Shane}, {Smith}, {Sobeck}, {Troup}, {Zamora},
  {Weinberg}, {Bovy}, {Eisenstein}, {Feuillet}, {Frinchaboy}, {Hayden},
  {Hearty}, {Nguyen}, {O'Connell}, {Pinsonneault}, {Wilson}, \&
  {Zasowski}}]{GarciaPerez2016}
{Garc{\'{\i}}a P{\'e}rez}, A.~E., {Allende Prieto}, C., {Holtzman}, J.~A.,
  {et~al.} 2016, \aj, 151, 144

\bibitem[{{Girardi}(1999)}]{Girardi1999}
{Girardi}, L. 1999, \mnras, 308, 818

\bibitem[{{Gunn} {et~al.}(2006){Gunn}, {Siegmund}, {Mannery}, {Owen}, {Hull},
  {Leger}, {Carey}, {Knapp}, {York}, {Boroski}, {Kent}, {Lupton}, {Rockosi},
  {Evans}, {Waddell}, {Anderson}, {Annis}, {Barentine}, {Bartoszek}, {Bastian},
  {Bracker}, {Brewington}, {Briegel}, {Brinkmann}, {Brown}, {Carr},
  {Czarapata}, {Drennan}, {Dombeck}, {Federwitz}, {Gillespie}, {Gonzales},
  {Hansen}, {Harvanek}, {Hayes}, {Jordan}, {Kinney}, {Klaene}, {Kleinman},
  {Kron}, {Kresinski}, {Lee}, {Limmongkol}, {Lindenmeyer}, {Long}, {Loomis},
  {McGehee}, {Mantsch}, {Neilsen}, {Neswold}, {Newman}, {Nitta}, {Peoples},
  {Pier}, {Prieto}, {Prosapio}, {Rivetta}, {Schneider}, {Snedden}, \&
  {Wang}}]{Gunn2006}
{Gunn}, J.~E., {Siegmund}, W.~A., {Mannery}, E.~J., {et~al.} 2006, \aj, 131,
  2332

\bibitem[{{Hekker} {et~al.}(2008){Hekker}, {Snellen}, {Aerts}, {Quirrenbach},
  {Reffert}, \& {Mitchell}}]{Hekker2008}
{Hekker}, S., {Snellen}, I.~A.~G., {Aerts}, C., {et~al.} 2008, \aap, 480, 215

\bibitem[{{Hekker} {et~al.}(2012){Hekker}, {Elsworth}, {Mosser}, {Kallinger},
  {Chaplin}, {De Ridder}, {Garc{\'{\i}}a}, {Stello}, {Clarke}, {Hall}, \&
  {Ibrahim}}]{Hekker2012}
{Hekker}, S., {Elsworth}, Y., {Mosser}, B., {et~al.} 2012, \aap, 544, A90

\bibitem[{{Holtzman} {et~al.}(2018){Holtzman}, {Hasselquist}, {Shetrone},
  {Cunha}, {Allende Prieto}, {Anguiano}, {Bizyaev}, {Bovy}, {Casey},
  {Edvardsson}, {Johnson}, {J{\"o}nsson}, {Meszaros}, {Smith}, {Sobeck},
  {Zamora}, {Chojnowski}, {Fernandez-Trincado}, {Garcia-Hernandez}, {Majewski},
  {Pinsonneault}, {Souto}, {Stringfellow}, {Tayar}, {Troup}, \&
  {Zasowski}}]{Holtzman:2018}
{Holtzman}, J.~A., {Hasselquist}, S., {Shetrone}, M., {et~al.} 2018, \aj, 156,
  125

\bibitem[{{Huber} {et~al.}(2009){Huber}, {Stello}, {Bedding}, {Chaplin},
  {Arentoft}, {Quirion}, \& {Kjeldsen}}]{Huber2009}
{Huber}, D., {Stello}, D., {Bedding}, T.~R., {et~al.} 2009, Communications in
  Asteroseismology, 160, 74

\bibitem[{{Huber} {et~al.}(2017){Huber}, {Zinn}, {Bojsen-Hansen},
  {Pinsonneault}, {Sahlholdt}, {Serenelli}, {Silva Aguirre}, {Stassun},
  {Stello}, {Tayar}, {Bastien}, {Bedding}, {Buchhave}, {Chaplin}, {Davies},
  {Garc{\'{\i}}a}, {Latham}, {Mathur}, {Mosser}, \& {Sharma}}]{Huber:2017}
{Huber}, D., {Zinn}, J., {Bojsen-Hansen}, M., {et~al.} 2017, \apj, 844, 102

\bibitem[{{Isaacson} \& {Fischer}(2010)}]{Isaacson:2010}
{Isaacson}, H., \& {Fischer}, D. 2010, \apj, 725, 875

\bibitem[{{Johns} {et~al.}(2018){Johns}, {Marti}, {Huff}, {McCann},
  {Wittenmyer}, {Horner}, \& {Wright}}]{Johns:2018}
{Johns}, D., {Marti}, C., {Huff}, M., {et~al.} 2018, ArXiv e-prints,
  arXiv:1808.04533

\bibitem[{{Kallinger} {et~al.}(2016){Kallinger}, {Hekker}, {Garcia}, {Huber},
  \& {Matthews}}]{Kallinger2016}
{Kallinger}, T., {Hekker}, S., {Garcia}, R.~A., {Huber}, D., \& {Matthews},
  J.~M. 2016, Science Advances, 2, 1500654

\bibitem[{{Lovis} \& {Fischer}(2010)}]{LovisFischer2010}
{Lovis}, C., \& {Fischer}, D. 2010, {Radial Velocity Techniques for
  Exoplanets}, ed. S.~{Seager}, 27--53

\bibitem[{{Majewski} {et~al.}(2017){Majewski}, {Schiavon}, {Frinchaboy},
  {Allende Prieto}, {Barkhouser}, {Bizyaev}, {Blank}, {Brunner}, {Burton},
  {Carrera}, {Chojnowski}, {Cunha}, {Epstein}, {Fitzgerald}, {Garc{\'{\i}}a
  P{\'e}rez}, {Hearty}, {Henderson}, {Holtzman}, {Johnson}, {Lam}, {Lawler},
  {Maseman}, {M{\'e}sz{\'a}ros}, {Nelson}, {Nguyen}, {Nidever}, {Pinsonneault},
  {Shetrone}, {Smee}, {Smith}, {Stolberg}, {Skrutskie}, {Walker}, {Wilson},
  {Zasowski}, {Anders}, {Basu}, {Beland}, {Blanton}, {Bovy}, {Brownstein},
  {Carlberg}, {Chaplin}, {Chiappini}, {Eisenstein}, {Elsworth}, {Feuillet},
  {Fleming}, {Galbraith-Frew}, {Garc{\'{\i}}a}, {Garc{\'{\i}}a-Hern{\'a}ndez},
  {Gillespie}, {Girardi}, {Gunn}, {Hasselquist}, {Hayden}, {Hekker}, {Ivans},
  {Kinemuchi}, {Klaene}, {Mahadevan}, {Mathur}, {Mosser}, {Muna}, {Munn},
  {Nichol}, {O'Connell}, {Parejko}, {Robin}, {Rocha-Pinto}, {Schultheis},
  {Serenelli}, {Shane}, {Silva Aguirre}, {Sobeck}, {Thompson}, {Troup},
  {Weinberg}, \& {Zamora}}]{Majewski2017}
{Majewski}, S.~R., {Schiavon}, R.~P., {Frinchaboy}, P.~M., {et~al.} 2017, \aj,
  154, 94

\bibitem[{{Ness} {et~al.}(2018){Ness}, {Silva Aguirre}, {Lund}, {Cantiello},
  {Foreman-Mackey}, {Hogg}, \& {Angus}}]{Ness2018}
{Ness}, M.~K., {Silva Aguirre}, V., {Lund}, M.~N., {et~al.} 2018, ArXiv
  e-prints, arXiv:1805.04519

\bibitem[{{Nidever} {et~al.}(2015){Nidever}, {Holtzman}, {Allende Prieto},
  {Beland}, {Bender}, {Bizyaev}, {Burton}, {Desphande}, {Fleming},
  {Garc{\'{\i}}a P{\'e}rez}, {Hearty}, {Majewski}, {M{\'e}sz{\'a}ros}, {Muna},
  {Nguyen}, {Schiavon}, {Shetrone}, {Skrutskie}, {Sobeck}, \&
  {Wilson}}]{Nidever2015}
{Nidever}, D.~L., {Holtzman}, J.~A., {Allende Prieto}, C., {et~al.} 2015, \aj,
  150, 173

\bibitem[{{Oelkers} \& {Stassun}(2018)}]{Oelkers_FFI:2018}
{Oelkers}, R.~J., \& {Stassun}, K.~G. 2018, \aj, 156, 132

\bibitem[{{Oelkers} {et~al.}(2018){Oelkers}, {Rodriguez}, {Stassun}, {Pepper},
  {Somers}, {Kafka}, {Stevens}, {Beatty}, {Siverd}, {Lund}, {Kuhn}, {James}, \&
  {Gaudi}}]{Oelkers:2018}
{Oelkers}, R.~J., {Rodriguez}, J.~E., {Stassun}, K.~G., {et~al.} 2018, \aj,
  155, 39

\bibitem[{{Oshagh} {et~al.}(2017){Oshagh}, {Santos}, {Figueira}, {Barros},
  {Donati}, {Adibekyan}, {Faria}, {Watson}, {Cegla}, {Dumusque}, {H{\'e}brard},
  {Demangeon}, {Dreizler}, {Boisse}, {Deleuil}, {Bonfils}, {Pepe}, \&
  {Udry}}]{Oshagh:2017}
{Oshagh}, M., {Santos}, N.~C., {Figueira}, P., {et~al.} 2017, \aap, 606, A107

\bibitem[{{Pande} {et~al.}(2018){Pande}, {Bedding}, {Huber}, \&
  {Kjeldsen}}]{Pande2018}
{Pande}, D., {Bedding}, T.~R., {Huber}, D., \& {Kjeldsen}, H. 2018, \mnras,
  480, 467

\bibitem[{{Pepper} {et~al.}(2012){Pepper}, {Kuhn}, {Siverd}, {James}, \&
  {Stassun}}]{Pepper:2012}
{Pepper}, J., {Kuhn}, R.~B., {Siverd}, R., {James}, D., \& {Stassun}, K. 2012,
  \pasp, 124, 230

\bibitem[{{Pepper} {et~al.}(2007){Pepper}, {Pogge}, {DePoy}, {Marshall},
  {Stanek}, {Stutz}, {Poindexter}, {Siverd}, {O'Brien}, {Trueblood}, \&
  {Trueblood}}]{Pepper:2007}
{Pepper}, J., {Pogge}, R.~W., {DePoy}, D.~L., {et~al.} 2007, \pasp, 119, 923

\bibitem[{{Pinsonneault} {et~al.}(2018){Pinsonneault}, {Elsworth}, {Tayar},
  {Serenelli}, {Stello}, {Zinn}, {Mathur}, {Garc{\'{\i}}a}, {Johnson},
  {Hekker}, {Huber}, {Kallinger}, {M{\'e}sz{\'a}ros}, {Mosser}, {Stassun},
  {Girardi}, {Rodrigues}, {Silva Aguirre}, {An}, {Basu}, {Chaplin}, {Corsaro},
  {Cunha}, {Garc{\'{\i}}a-Hern{\'a}ndez}, {Holtzman}, {J{\"o}nsson},
  {Shetrone}, {Smith}, {Sobeck}, {Stringfellow}, {Zamora}, {Beers},
  {Fern{\'a}ndez-Trincado}, {Frinchaboy}, {Hearty}, \&
  {Nitschelm}}]{Pinsonneault2018}
{Pinsonneault}, M.~H., {Elsworth}, Y.~P., {Tayar}, J., {et~al.} 2018, ArXiv
  e-prints, arXiv:1804.09983

\bibitem[{{Pont} {et~al.}(2011){Pont}, {Aigrain}, \& {Zucker}}]{Pont:2011}
{Pont}, F., {Aigrain}, S., \& {Zucker}, S. 2011, \mnras, 411, 1953

\bibitem[{{Ricker} {et~al.}(2015){Ricker}, {Winn}, {Vanderspek}, {Latham},
  {Bakos}, {Bean}, {Berta-Thompson}, {Brown}, {Buchhave}, {Butler}, {Butler},
  {Chaplin}, {Charbonneau}, {Christensen-Dalsgaard}, {Clampin}, {Deming},
  {Doty}, {De Lee}, {Dressing}, {Dunham}, {Endl}, {Fressin}, {Ge}, {Henning},
  {Holman}, {Howard}, {Ida}, {Jenkins}, {Jernigan}, {Johnson}, {Kaltenegger},
  {Kawai}, {Kjeldsen}, {Laughlin}, {Levine}, {Lin}, {Lissauer}, {MacQueen},
  {Marcy}, {McCullough}, {Morton}, {Narita}, {Paegert}, {Palle}, {Pepe},
  {Pepper}, {Quirrenbach}, {Rinehart}, {Sasselov}, {Sato}, {Seager},
  {Sozzetti}, {Stassun}, {Sullivan}, {Szentgyorgyi}, {Torres}, {Udry}, \&
  {Villasenor}}]{Ricker:2015}
{Ricker}, G.~R., {Winn}, J.~N., {Vanderspek}, R., {et~al.} 2015, Journal of
  Astronomical Telescopes, Instruments, and Systems, 1, 014003

\bibitem[{{Saar}(2003)}]{Saar:2003}
{Saar}, S.~H. 2003, in Astronomical Society of the Pacific Conference Series,
  Vol. 294, Scientific Frontiers in Research on Extrasolar Planets, ed.
  D.~{Deming} \& S.~{Seager}, 65--70

\bibitem[{{Saar} {et~al.}(1998){Saar}, {Butler}, \& {Marcy}}]{Saar:1998}
{Saar}, S.~H., {Butler}, R.~P., \& {Marcy}, G.~W. 1998, \apjl, 498, L153

\bibitem[{{SDSS Collaboration} {et~al.}(2016){SDSS Collaboration}, {Albareti},
  {Allende Prieto}, {Almeida}, {Anders}, {Anderson}, {Andrews},
  {Aragon-Salamanca}, {Argudo-Fernandez}, {Armengaud}, \& et~al.}]{DR13}
{SDSS Collaboration}, {Albareti}, F.~D., {Allende Prieto}, C., {et~al.} 2016,
  ArXiv e-prints, arXiv:1608.02013

\bibitem[{{Serenelli} {et~al.}(2017){Serenelli}, {Johnson}, {Huber},
  {Pinsonneault}, {Ball}, {Tayar}, {Silva Aguirre}, {Basu}, {Troup}, {Hekker},
  {Kallinger}, {Stello}, {Davies}, {Lund}, {Mathur}, {Mosser}, {Stassun},
  {Chaplin}, {Elsworth}, {Garcia}, {Handberg}, {Holtzman}, {Hearty},
  {Garcia-Hernandez}, {Gaulme}, \& {Zamora}}]{Serenelli:2017}
{Serenelli}, A., {Johnson}, J., {Huber}, D., {et~al.} 2017, ArXiv e-prints,
  arXiv:1710.06858

\bibitem[{{Stassun} {et~al.}(2017){Stassun}, {Collins}, \&
  {Gaudi}}]{Stassun:2017}
{Stassun}, K.~G., {Collins}, K.~A., \& {Gaudi}, B.~S. 2017, \aj, 153, 136

\bibitem[{{Stassun} {et~al.}(2018{\natexlab{a}}){Stassun}, {Corsaro}, {Pepper},
  \& {Gaudi}}]{Stassun:2018}
{Stassun}, K.~G., {Corsaro}, E., {Pepper}, J.~A., \& {Gaudi}, B.~S.
  2018{\natexlab{a}}, \aj, 155, 22

\bibitem[{{Stassun} {et~al.}(2018{\natexlab{b}}){Stassun}, {Oelkers}, {Pepper},
  {Paegert}, {De Lee}, {Torres}, {Latham}, {Charpinet}, {Dressing}, {Huber},
  {Kane}, {L{\'e}pine}, {Mann}, {Muirhead}, {Rojas-Ayala}, {Silvotti},
  {Fleming}, {Levine}, \& {Plavchan}}]{Stassun_TIC:2018}
{Stassun}, K.~G., {Oelkers}, R.~J., {Pepper}, J., {et~al.} 2018{\natexlab{b}},
  \aj, 156, 102

\bibitem[{{Walker} {et~al.}(2003){Walker}, {Matthews}, {Kuschnig}, {Johnson},
  {Rucinski}, {Pazder}, {Burley}, {Walker}, {Skaret}, {Zee}, {Grocott},
  {Carroll}, {Sinclair}, {Sturgeon}, \& {Harron}}]{Walker:2003}
{Walker}, G., {Matthews}, J., {Kuschnig}, R., {et~al.} 2003, PASP, 115, 1023

\bibitem[{{Wright}(2005)}]{Wright:2005}
{Wright}, J.~T. 2005, \pasp, 117, 657

\bibitem[{{Yu} {et~al.}(2018){Yu}, {Huber}, {Bedding}, \& {Stello}}]{Yu:2018}
{Yu}, J., {Huber}, D., {Bedding}, T.~R., \& {Stello}, D. 2018, \mnras, 480, L48

\end{thebibliography}

\end{document}